\documentclass[12pt,preprint]{aastex}

\slugcomment{To appear in ApJ}

\shorttitle{Dust around RCB stars}
\shortauthors{Garc\'{\i}a-Hern\'andez et al.}

\begin{document}


\title{Dust around R Coronae Borealis stars: I. Spitzer/IRS observations}


\author{D. A. Garc\'\i a-Hern\'andez\altaffilmark{1,2}, N. Kameswara
Rao\altaffilmark{3,4}, David L. Lambert\altaffilmark{4}}


\altaffiltext{1}{Instituto de Astrof\'{\i}sica de Canarias, C/ Via L\'actea
s/n, 38200 La Laguna, Spain; agarcia@iac.es}
\altaffiltext{2}{Departamento de Astrof\'{\i}sica, Universidad de La Laguna (ULL), E-38205 La Laguna, Spain}
\altaffiltext{3}{Indian Institute of Astrophysics, Bangalore 560034, India; nkrao@iiap.res.in}
\altaffiltext{4}{W. J. McDonald Observatory. The University of Texas at
Austin. 1 University Station, C1400. Austin, TX 78712$-$0259, USA; dll@astro.as.utexas.edu}


\begin{abstract}
Spitzer/IRS spectra from 5 to 37 $\mu$m for a complete sample of 31 R Coronae
Borealis stars (RCBs) are presented. These spectra are combined with optical
and near-infrared photometry of each RCB at maximum light to compile a spectral
energy distribution (SED). The SEDs are fitted with blackbody flux
distributions and estimates made of the ratio of the infrared flux from
circumstellar dust to the flux emitted by the star. Comparisons for 29 of the
31 stars are made with the IRAS fluxes from three decades earlier: Spitzer and
IRAS fluxes at 12 $\mu$m and 25 $\mu$m are essentially  equal for all but a
minority of the sample.  For this minority, the IRAS to Spitzer flux ratio
exceeds a factor of three. The outliers are suggested to be  stars where
formation of a dust cloud or dust puff is a rare event. A single puff  ejected
prior to the IRAS observations may have been reobserved by Spitzer  as a cooler
puff at a greater distance from the RCB. RCBs which experience more frequent
optical declines have, in general, a circumstellar environment containing puffs
subtending a larger solid angle at the star and a quasi-constant infrared flux.
Yet, the estimated subtended solid angles and the blackbody temperatures of the
dust show a systematic evolution to lower solid angles and cooler temperatures
in the interval between IRAS and Spitzer. Dust emission by these RCBs and
those in the LMC is similar in terms of total 24 $\mu$m luminosity and
[8.0]$-$[24.0] color index.
\end{abstract}


\keywords{circumstellar matter --- dust, extinction --- stars: chemically
peculiar --- stars: white dwarfs --- infrared: stars}



\section{Introduction}

The R Coronae Borealis (here, RCB) stars are notable for two distinct peculiarities
(Clayton 1996). First, they are hydrogen-poor, helium-rich supergiants: the
H-deficiencies range from about 10-100 to at least 10$^8$. Second, the RCB stars
experience unpredictable and rapid declines in brightness: declines of 2 to 8
magnitudes in the visual occuring at intervals of less than a year to greater than
20 years last from weeks to months to years. These declines are caused by formation
of a cloud of carbon soot above the Earth-facing surface of the star. Discovery of
an infrared excess confirmed the obvious suspicion that the stars were dust
producers (Stein et al. 1969; Feast et al. 1997). Typical blackbody temperatures of
the dust run from about 400 K to 900 K. In a representative case, about one-third
of the photospheric flux is absorbed by dust and reemitted in the infrared. The
observation that the infrared flux may be little affected by a decline  shows that
the dust is distributed in clouds around the star (Forrest et al. 1972).  Recently,
high angular resolution images of RY Sgr, a bright RCB, at 2.2, 4.5, and
8$-$13 $\mu$m showed clearly that the dust is indeed distributed in clouds (de
Laverny \& M\'{e}karnia 2004; Le\~ao et al. 2007; Bright et al. 2011). IRAS
photometry  at long wavelegnths showed that, in addition to the warm dust, some
RCBs have dust at a lower temperature (say, 30 K to 100 K) and, therefore, at large
distances from the star (Rao \& Nandy 1986; Walker 1986; Gillett et al. 1986). 

Each of the two principal peculiarities prompts  leading questions. In the case
of the H-deficiency, that question is - what are the evolutionary origins of the
RCBs that result in a H-poor stars?  Two scenarios remain under active
consideration for the RCBs and their putative relatives the H-deficient carbon
(HdC) stars to lower temperatures and the extreme helium (EHe) stars to higher
temperature. In one, the H-poor supergiant is formed from the merger of a He
white dwarf with a C-O white dwarf; the double-degenerate (DD) scenario. In the
competing picture, the H-poor supergiant results from a final post-AGB shell
flash in the central star of a planetary nebula; the so-called final flash (FF)
scenario. In both cases, the trigger - the merger or the final flash -
transforms a white dwarf into a H-poor supergiant  for a period of a few
thousand years. There is evidence that both the DD and FF scenarios occur but
the DD scenario seems  likely to account for the majority of the RCBs. 

Several insights into the origins of the RCBs are coming from spectroscopic
determinations of the stellar chemical compositions (Lambert \& Rao 1994;
Asplund et al. 2000; Clayton et al. 2005, 2007; Garc\'{\i}a-Hern\'{a}ndez et al.
2009, 2010; Jeffery et al. 2011; Pandey \& Lambert 2011). Detailed
abundance analyses, which are possible for the warm RCBs but rarely undertaken
for the cool RCBs with their spectra rich in molecular lines, suggest that many
RCBs are  likely fruits of the DD scenario. A few RCBs show several highly
unusual abundance signatures and, in particular, very distinctive Si/Fe and S/Fe
ratios. Such stars are called `minority' RCBs -  see Lambert \& Rao (1994) who
introduced the terms `majority RCB' and `minority RCB' star.  A rare class of
hot RCBs is discussed by De Marco et al. (2002) and includes DY Cen, a minority
RCB.

The other principal peculiarity -- the unpredictable declines -- stimulates a
series of  questions about the dust around RCB stars such as: What is the
composition of the dust? Where does dust form relative to the stellar surface?
What triggers  formation of the obscuring cloud?  How frequently do dust clouds
form? Does dust form at preferred latitudes on the star or are formation sites
spread uniformly across the stellar surface?  Clayton (1996) reviews evidence
pertinent to these  questions. Perhaps, the key novel theoretical idea of recent
times comes from Woitke et al. (1996) who developed a model in which a
pulsation-induced shock triggers dust nucleation near the star (one to two
stellar radii out) as  gas behind the outward propagating shock cools below the
condensation temperature (say, $< 1500$ K) in the star's upper atmosphere.
Presence of cool gas during light minima has now been detected in three RCBs: R
CrB (Rao et al. 2006; Rao \& Lambert 2010), V854 Cen (Rao \& Lambert 2000), and
V CrA (Rao \& Lambert 2008).  Light variations at maximum light are common among
RCBs and generally interpreted as arising from pulsations. Absorption line
splitting suggestive of an atmospheric shock is regularly observed for RY Sgr
(Danziger 1965; Cottrell \& Lambert 1982) and occasionally for R CrB. Pugach
(1977) noted a correlation between the onset of a light decline for RY Sgr  and
the pulsation phase. Crause et al.  (2007) from long-term photometric studies of
four RCBs have shown that a decline occurs at a particular phase of the
pulsation cycle, although not every pulsation cycle results in a decline. Thus,
evidence is accumulating that dust formation occurs near the star.  Radiation
pressure on the dust grains is considered to drive them them rapidly outward.

Clues to several of the questions concerning the dust are contained in the shape
of the infrared continuum emitted by the circumstellar dust, the presence of
emission or absorption features imposed on that continuum, and on the temporal
variation of  the infrared flux.  Ground-based infrared spectrophotometry has
revealed a smooth continuum in the atmospheric windows; strong  emission and
absorption features have not been seen. Inability to observe in regions blocked
by the Earth's atmosphere is an especially serious problem in searching for
features attributable to dust.  Infrared Space Observatory (ISO) spectra
obtained for just three (the brightest) RCBs - R CrB, RY Sgr, and V854 Cen - at
a resolution of R=1000 showed for the first time some excess emission over a
quasi-blackbody continuum (Lambert et al. 2001). Broad  unidentified emission
features were seen centered on about 6 $\mu$m and 13 $\mu$m. Emission features
from 4 $\mu$m to 15 $\mu$m for V854 Cen but not for R CrB and RY Sgr showed a
resemblance to a laboratory spectrum of hydrogenated amorphous carbon (Colangeli
et al. 1995; Scott et al. 1997). 

With the advent of the Spitzer Space telescope, it became possible for the
first time to extend  low-resolution infrared spectroscopy to a much larger
sample of RCBs. In this paper, we present a library of infrared Spitzer/IRS
spectra for a large sample (31) of RCB stars. Spectra are characterized by
a fit of blackbodies to optical and  infrared photometry and the Spitzer
spectra. Estimates are provided of the infrared flux emitted by dust to the
total stellar flux. Comparisons are made with previously reported measurements
of the infrared flux, principally the 12$\mu$m and 25$\mu$m flux measurements 
from the IRAS satellite. Infrared emission features superimposed on the
blackbody continuua will be discussed in detail in a subsequent paper. Emission
spectra of DY Cen and V854 Cen showing features from polycyclic aromatic
hydrocarbons (PAHs) and C$_{60}$ have been discussed by
Garc\'{\i}a-Hern\'{a}ndez et al. (2011). Spitzer/IRS spectra of the hot RCB
stars V348 Sgr and HV 2671 are presented in Clayton et al. (2011).

Section 2 describes our sample of RCB stars, the Spitzer/IRS and  some ground-based
photometric observations obtained at about  the same time as the Spitzer
observations, and the IRAS observations.  Section 3 constructs the  spectral energy
distributions (SEDs) from $\sim$0.4 to 38 $\mu$m  from optical and infrared
observations, where the SEDs are fitted using blackbodies for the star and the dust
for each object. Spitzer and IRAS observations are compared and discussed in
Section 4, while a comparison of the RCB dust emission in different
metallicity environments is offered in Section 5. Finally, the paper concludes with
Section 6. 

\section{The sample and observations}

\subsection{The RCB sample}

Our main goal in obtaining Spitzer observations was to compile a library of
infrared spectra for as complete a sample of RCBs as possible.   Table 1 list
the 31 RCB stars included in this study together with some relevant information
such as coordinates, the  date of the Spitzer observation, Spitzer program ID,
etc.  Eighteen RCBs were in our approved GO program. An additional 13 warm RCBs
were found in the public Spitzer database. The sample provides comprehensive
coverage of the hot RCBs, warm RCBs and includes several of the coolest RCBs.
The target list in Table 1 is also identified according to  the categories A, B,
or C.  Category A corresponds to  warm RCBs across the composition range (13
stars) where compositions are taken from Asplund et al. (2000).  Category B is
assigned to the few (5) minority RCB stars. The coolest RCB stars belong to
category C (13 stars). Note that minority stars V3795 Sgr, VZ Sgr, V CrA, V854
Cen and DY Cen fall in categories A and B (AB) and Z UMi is assigned to the
category BC. The RCB star HV 2671 in the Large Magellanic Cloud (LMC) has yet to
be assigned to one of the categories; De Marco et al. (2002) note that HV 2671
and V348 Sgr have almost identical optical spectra but the Spitzer spectra
are very different although HV 2671 shows similarities with V854 Cen (Clayton et
al. 2011).  Also in Table 1, we indicate whether a star was observed at or
below maximum light.

\subsection{Spitzer observations}

The infrared spectra were taken with the Infrared Spectrograph (IRS, Houck et
al. 2004) on board the Spitzer Space Telescope (Werner et al. 2004). We obtained
5.2$-$37.2 $\mu$m spectra for 18 sources in our sample under our General
Observer Program (\#50212, P.I.: D. L. Lambert) that was carried out between
April and October 2008. We used a combination of IRS Short-Low (5.2$-$14.5
$\mu$m; 64 $<$ R $<$ 128, here SL), Short-High (9.9$-$19.6 $\mu$m, here SH)  and
Long-High (18.7$-$37.2 $\mu$m, here LH) observations (R$\sim$600). Since IRAS
fluxes at 12 and 25 $\mu$m  are available for all sources in our sample, we
assumed we had {\it a priori}  knowledge of the mean brightness of each source
at the different wavelengths covered by IRS.   Most of these sources are very
bright (with mid-IR SEDs peaking at $\sim$12 $\mu$m), and two cycles of 6 s in
each of the three modules were used. For those sources with lower flux densities
at 25 $\mu$m, four cycles of 14 s were employed in the LH module. We typically
reached a S/N larger than 50 in the SL and SH modules; these modules cover the
5.2$-$19.5 $\mu$m range where most of the spectral features of our interest
fall. However, the achieved S/N is generally lower in the LH module. For three
stars (those sources brighter than 5.5 Jy at 12 $\mu$m; see Table 1) we did not
obtain spectroscopy in the SL module in order to avoid saturation.

IRS spectra for 13 other stars were retrieved from the Spitzer database. These
spectra were taken by different observers and using different module
combinations (see Table 1). In general, the quality of these spectra is also
very good (S/N$\geq$50); specially when the SL and LL (Long-Low: 14.0$-$38.0
$\mu$m; 64 $<$ R $<$ 128) modules are used. The Spitzer/IRS spectra of the
infrared-bright RCB stars V854 Cen, RY Sgr and R CrB - previously observed by
the ISO satellite (Lambert et al. 2001) - are included in this subgroup. 

We retrieved the 1-D infrared spectra processed by the Spitzer data reduction
pipeline (versions 15.3.0, 16.1.0, 17.2.0, 18.0.2 and 18.7.0) for all sources in
our sample from the Spitzer database. These post-bcd products (one spectra for
each nod position) are automatically reduced by the IRS Custom Extractor (SPICE)
with a point source aperture. The automatic data reduction includes the
extraction from the 2-D images as well as the wavelength and flux calibration.
It is to be noted here that for the SL and LL data, the two nod position 2-D
images are subtracted in order to cancel out the sky background. However, for
the high-resolution modules no background subtraction is done since no sky
measurements were taken; the SH and LH slits are too small for on-slit
background subtraction. The Spitzer-contributed software SMART (Higdon et al.
2004) was later used for cleaning of residual bad pixels, spurious jumps and
glitches and for smoothing and merging into one final spectrum per source. Note
that all sources in our sample are bright point-like objects for which the
automatic data reduction pipeline works very well; no significant differences
between these spectra and those manually reduced are found.

We found a good match (i.e., better than 5\%) between the different modules for
approximately half of the sample stars. Most of the rest of stars displayed a
very good match between the SL and SH modules, confirming their point-like
nature but the LH data showed a flux excess of about 5$-$20\%. We attribute this
mismatch to the fact the these LH fluxes are more uncertain and to possible
background emission. Indeed, several of these stars are located toward high
extinction line of sights (i.e., their infrared spectra are affected by
amorphous silicate absorptions from the diffuse interstellar medium; see below).
Thus, we applied a correction factor to the LH observations in order to scale
them to the SH spectra. It should be noted that only the RCB star VZ Sgr  seems
to be slightly extended at infrared wavelengths, 

Reduced spectra -- $\lambda F_\lambda$ versus $\lambda$ -- are shown in
Figure 1-5 for the complete sample of 31 stars. 

\subsection{Ground-based optical/near-IR photometry}

To complement the Spitzer observations, we carried out photometric observations
in the optical and near-infrared for some of the sample.  Our intention was to
ascertain a star's  status (i.e., if the stars were at maximum or minimum
light)  during the Spitzer observations. However, these ``simultaneous"
photometric data were obtained only for the 18 RCB stars observed with Spitzer
through Program 50212 (Table 2).  If a star was at maximum light, the photometry
is used in the  construction of the spectral energy distribution (SEDs) from the
visible to $\sim$40 $\mu$m in our sample stars. For stars not at maximum light,
photometry from the literature was used to establish the stellar energy
distribution at maximum light across the optical.

Optical photometry in the Johnson-Bessell V, R, and I filters was obtained with the
IAC-80 telescope (Observatorio del Teide, Spain) equipped with the CAMELOT
CCD\footnote{see e.g.,
http://www.iac.es/telescopes/pages/es/inicio/instrumentos/camelot.php} for more
details. The observations were done as near as possible to the Spitzer observation
dates and sometimes the stars were observed twice (i.e., before and after Spitzer). The
VRI magnitudes for each star were derived by using standard aperture photometry
tasks in IRAF\footnote{Image Reduction and Analysis Facility (IRAF) software is
distributed by the National Optical Astronomy Observatories, which is operated by the
Association of Universities for Research in Astronomy, Inc., under cooperative agreement
with the National Science Foundation.}. The flux calibration was done by using the
photometric calibration for CAMELOT\footnote{see
http://www.iac.es/telescopes/pages/en/home/utilities.php\#camelot-calibration for the
average extinction coefficients, color terms, etc.} and making use of standard stars
observed on the same night. The use of this average photometric calibration implies that
our derived VRI magnitudes are precise to $\sim$0.15 mag. This error in the optical
magnitudes is more than enough for our purposes, that is, to know the variability status
of these RCB stars. Table 2 displays a log of the optical observations done (e.g.,
the observation dates) together with the VRI photometry for each star observed.

JHKL photometry (Table 2) was obtained at the South African Astrophysical
Observatory (SAAO) with the 0.75m telescope by F. Van Wyk at our request. These
observations are on the SAAO system using Carter (1990) standards. 

\section{Spectral Energy Distributions}

\subsection{Methodology}

In this section, we present the  spectral energy distribution (SED) from
$\sim$0.4 to 40 $\mu$m for each RCB star.  UBVRIJHKLMN magnitudes  are converted
to fluxes with the magnitude-flux and effective wavelength calibrations taken
from Tokunaga (2000). For a majority of the stars, there are UBVRIJHK 
magnitudes in the literature. Some data are available for L magnitudes and a few
for MN magnitudes. In addition, there are the ground-based measurements in Table
2.

The goal was to construct the `stellar' SED from observations made when the star
was at maximum light. This SED uses the UBVRIJHK fluxes except that for a few
stars some observations of  K and possibly H are contaminated by emission by
dust.  In addition to the Spitzer spectrum of the dust emission,  we consider
the IRAS 12$\mu$m and 25$\mu$m measurements and  available LMN photometry.  The
LMN, Spitzer and IRAS fluxes are primarily from the circumstellar dust.  As is
well known, optical and infrared variability are not tightly coupled - see below
and especially Feast et al. (1997).

These SEDs are corrected for interstellar reddening provided by the line of
sight to the RCB. Spitzer spectra  require a correction for  absorptions at
9.7$\mu$m and 18$\mu$m, attributable to interstellar silicates. For this
correction, the reddening curve is adopted from Chiar \& Tielens (2006) by
taking A(K)/A(V)=0.114 (Cardelli et al. 1989) with an extrapolation from 27 to
38 $\mu$m assuming the same slope as between 23 to 27 $\mu$m. The correction was
generally ignored for stars where the predicted reddening was less than about
E(B-V) of 0.4. The correction for the 9.7 $\mu$m interstellar absorption can
have a particularly strong effect on the profile and intensity of the 6-10
$\mu$m emission feature, the subject of a subsequent paper.  

A few stars in Table 1 were observed by Spitzer in decline. For these stars, we
assemble the stellar SED from published photometry at maximum light; we do not
use the contemporary photometry, if available, in Table 2.  This is then
combined with the Spitzer spectrum to provide that star's SED which is
corrected, as usual, for interstellar reddening.

In the case of the IRAS 12 $\mu$m and 25 $\mu$m photometry, the measurements
were color corrected. Color-corrected flux densities were mostly obtained from
Walker (1986). In some cases, measurements in the IRAS Point Source Catalogue
were corrected following the prescriptions given by Beichman et al.  (1988 -
Table VI.C.6).

Each reddening-corrected  SED was fitted  with a combination of 
blackbodies with one blackbody at the stellar effective temperature and  one or
two blackbodies to represent the infrared circumstellar component. In a few
cases, a third circumstellar blackbody was considered.  Table 3 summarizes the
fits by giving the temperature and the estimated flux ratio for the blackbody
relative to the stellar flux where $R=f_{cool}/f_{star}$ is referred to as the
covering factor.  The sum of the $R$-values for a given star is essentially
independent of the assumption that dust emission may be represented by one or
more blackbodies. Our $R$-values do not include the small contribution from the
6-10 $\mu$m emission feature. Entries are given for both the Spitzer and IRAS
fits except where there is no significant difference between the two fits.

All  sources of photometry and interstellar reddening are identified below where
brief descriptions are also given of other charcteristics of each star
(frequency of declines, comparison with IRAS and other IR fluxes,  etc.). A
primary source on the frequency of declines is Jurcsik (1996) who compiled the
inter-fade periods for a majority of our sample. She defines a fading of a RCB
to be `an initial drop of about 1 mag from a maximum light, independently of the
duration and complexity of the minima.' The AAVSO website provides a historical
record of the light curves of many of our RCBs from which we also estimate the
frequency of declines. In addition, the ASAS-3\footnote{See
http://www.astrouw.edu.pl/asas/} website provides 9 year light curves in the
V-band for many stars in our sample, covering the epoch of the Spitzer
observations.

\subsection{Individual Stars}

{\bf UV Cas:}  UV Cas is very rarely seen in decline. Zavatti (1975) from a
sparse data set assembled from the literature found ``during 69 years of
observations only one deep minimum'', a minimum of about four magnitudes
recorded more than eighty years ago. An excellent data set from the AAVSO
extending back to the 1950s shows no deep declines in last 60 years. There is
evidence for  a 1.2 magnitude decline between 1954 August and 1956 August and
perhaps one or two even weaker declines but none for the past 30 years. Jurcsik
(1996) gives the inter-fade period as 25500 days, among her sample of 27 RCBs
only XX Cam at 36000 days fades less frequently. 

VRI photometry  (Table 2) was obtained at the time of the Spitzer observations. 
We take UBV photometry from Fernie et al. (1972); the star is only slightly
variable in UBV. 2MASS JHK magnitudes are adopted (Cutri et al. 2003).  A
valuable set of JHKLM photometry from 1984--2009 is provided by Bogdanov et al.
(2010). 

The Spitzer spectrum shows the interstellar 9.7 $\mu$m silicate absorption band
which is almost entirely removed when  the spectrum is corrected assuming E(B-V)
= 0.9.  Rao (1980) estimated E(B-V)=1.0 from interstellar reddening maps
(Fitzgerald 1968) and the assumption that this luminous star must be beyond the
majority of the reddening. 

Corrected for interstellar reddening, the UBVRI magnitudes are fitted by a 7200
K blackbody, close to the effective temperature found from optical spectroscopy
by Asplund et al. (2000). The 2MASS JHK fluxes imply a brighter blackbody by
about 0.2 magnitudes. The Spitzer fluxes are fitted with the  principal
contribution from a 510 K blackbody and minor contributions from the Planck
tail of the stellar blackbody and a colder (180 K) blackbody (Figure 6 --
lefthand panel, and Table 3).  The covering factor $R$ sums to 0.035 for the two
cool blackbodies (Table 3), one of the lowest $R$ for the entire sample. Also,
the 6-10$\mu$m excess emission is very weak and dependent on the correction for
interstellar extinction.

UV Cas shows a large flux variation between IRAS and Spitzer observations. The
IRAS 12 $\mu$m and 25 $\mu$m fluxes are factors of 6 (the highest value for our
sample) and 3 (the second highest value for our sample),  respectively, greater
than the Spitzer values. A fit to the IRAS fluxes and ground-based photometry
requires a blackbody at about 800 K and a higher $R$ (=0.28) than required by
the Spitzer fluxes (Figure 6 -- righthand panel).

Bogdanov et al.'s (2010) survey shows that at the time of the IRAS observations
UV Cas was unusually bright in the infrared. The L magnitude was almost two
magnitudes brighter than when the star was observed by Spitzer  and 0.8
magnitudes brighter than the 1973 measurement reported by Rao (1980). The M
magnitude was about 1.5 magnitudes brighter than in 2008.  This IR excess
observed by IRAS decayed over about 2000 days and was followed much later by two
minor increases by about 0.6 magnitudes in L for a duration of about 1000 days
without a pronounced optical decline. Evidently, UV Cas is an irregular infrared
variable without contemporary optical variability. Bogdanov et al.'s L and M
magnitudes are quite well reproduced by the fit to the IRAS fluxes (Figure  6 --
righthand panel).

{\bf S Aps:} S Aps is a cool RCB  with a slightly less than average tendency to
go into decline; Jurcsik (1996) gives the inter-fade period as 1400 days. 

The UBVRIJHKLMN magnitudes  were assembled from the following sources: UBV
(Zhilyaev et al. 1978), UBVRI (Marang et al. 1990),  JHKL (Table 2), MN
(Kilkenny \& Whittet 1984). The adopted reddening is E(B-V) = 0.05 (Asplund et
al. 1997).  Feast et al. (1997) estimated E(B-V) = 0.13 but  at such low
reddenings the correction to the Spitzer and IRAS fluxes is unimportant.

A blackbody fit to the dereddened optical photometry  gives a stellar
temperature of 4200 K. De-reddened Spitzer fluxes and the contemporaneous JHKL
photometry are well fit with the stellar 4200 K and a dust blackbody at 750 K with
dust dominating the star at  wavelengths at K and beyond. The covering factor
$R=0.37$ is a typical value. 

S Aps is a  striking example where not only are the IRAS and Spitzer fluxes 
very similar  but where  measures of the IR excess at other times  indicate an
almost invariant excess and suggest a circumstellar environment containing a
large number of dust clouds.  For example, earlier photometry at KL (Glass 1978;
Feast et al. 1997) and MN (Kilkenny \& Whittet 1984) are reproduced
satisfactorily by the stellar-dust blackbody combination. Variations of no more
than several tenths of a magnitude at L are indicative  of only minor variations
in the cloud  population in the circumstellar environment.

{\bf SV Sge:}    SV Sge is a cool RCB  experiencing declines at a typical rate;
Jurcsik (1996) gives the inter-fade period as 2500 days.   Spitzer observed the
star at maximum (Table 2). 

The UBVRI magnitudes for maximum light are taken as follows: VRI (Table 2),  JHK
(2MASS) and  a B magnitude  by assuming a (B-V) identical to  that of the HdC
star HD 137613  because the HdC and SV Sge have similar K-band spectra
(Garc\'{\i}a-Hern\'{a}ndez et al. 2010). A reddening E(B-V) = 0.72 is adopted; a
larger reddening results in an emission bump at 9.7 $\mu$m, a feature shown by
no other RCB.

The SED is well fit with a stellar blackbody of 4200 K and dust blackbodies of
565 K and 350 K  which in total correspond to a covering fraction $R=0.074$.  The
IRAS 25 $\mu$m flux agrees well with the Spitzer flux but the 12 $\mu$m flux is
almost double the Spitzer value which demands a hotter blackbody (720 K), a
larger covering factor $R=0.15$  with the same stellar blackbody. This flux
increase at the time of the IRAS measurement implies a rather fresh ejection of
dust. Perhaps, this ejection was responsible for the three magnitude optical
decline which began in 1981 January and ended 1982 November. The dust emission
contributes a few per cent of the K flux at the time of the Spitzer observation
but approaches 10 per cent at the time of the IRAS observations.  

{\bf Z UMi:}  This cool star  identified as a RCB by Benson et al. (1994) is
frequently in decline; since the star was put on the AAVSO  program, it has
shown nine declines in 16 years.  Kipper \& Klochkova's (2006) analysis led them
to suggest Z UMi is a minority RCB of low metallicity with a lithium excess
(Goswami et al. 1997).  Jurcsik (1996) did not include Z UMi in her
determination of inter-fade periods.  Spitzer observations were obtained as Z
UMi was about two magnitudes below maximum light  and  recovering from the
deepest longest lasting decline on record.   The star is at a Galactic latitude
of 33$^\circ$ and, therefore, we assume negligible interstellar reddening. 

The stellar  blackbody temperature is taken as 5200 K, a value consistent with
the spectroscopic estimate of 5250$\pm$150 K (Kipper \& Klochkova 2006). 
Photometry for BVI from AAVSO and JHK from 2MASS at maximum light  is fitted by
this blackbody. A fit to the Spitzer spectrum calls for a blackbody at 710 K and
a covering factor $R=0.43$.  Since the optical depth of the cloud(s) along the
line of sight likely varies with wavelength, the adopted fit underestimates the
(small) stellar contribution at infrared wavelengths. 

The IRAS fluxes about 60\% greater than Spitzer values suggest a blackbody at
850 K and a covering factor $R=0.95$, a value higher than from the Spitzer fluxes
at a time when the star was below maximum light.

{\bf V1783 Sgr:} At the time of the Spitzer observation, V1783 Sgr was  near
maximum light following the deepest decline in 20 years, a decline that began
about 2002 October and ended with restoration to maximum light about 2007 July.
Apart from this unusually long but not particularly deep (about three
magnitudes) decline, V1783 Sgr has shown only three  declines over two decades. 

Photometry obtained almost simultaneously with the Spitzer observations is in
Table 2.  Extensive UBVRI photometry at maximum light was reported by Lloyd
Evans et al. (1991) who proposed V1783 Sgr as a cool RCB. The VRI magnitudes  in
Table 2 are within the range reported by Lloyd Evans et al. An interstellar
E(B-V)= 0.42 is suggested by the elimination of the absorptions at 9.7 $\mu$m and
18 $\mu$m from the Spitzer spectrum.

By combining  Lloyd Evans et al.'s photometry with the JHK photometry from Table 2 
and correcting for the interstellar reddening,  the SED is fit with a stellar
blackbody temperature of 5600 K and dust blackbody of 560 K for a covering factor
of $R=0.28$. The color-corrected IRAS 12 $\mu$m and 25 $\mu$m fluxes are
similar to the Spitzer values, giving a sligthly hotter dust blackbody of 600 K with
a covering factor of $R=0.30$.

The 2MASS JHK are about a magnitude brighter than values in Table 2,  and since
the star was at visual maximum at the time of the 2MASS observation,  it would
appear that warm dust was present but off the line of sight.

{\bf WX CrA:} This cool RCB  observed by Spitzer at maximum light experiences
declines at a typical frequency; Jurcsik (1996) gives the inter-fade period as
2000 days.  Maximum light UBVRI photometry  was taken from Marang et al. 1990).
The 2MASS JHK photometry agrees well with earlier measurements by Feast et al.
(1997). Feast et al's L magnitude and Kilkenny \& Whittet's (1984) M and N
magnitude complete the available photometry at maximum light.  Interstellar
reddening is slight: E(B-V)= 0.06 (Rao 1995 in Asplund et al. 1997; Feast et al.
1997)

A fit to the maximum light reddening-corrected   photometry and the Spitzer 
spectrum calls for a stellar blackbody at 4200 K and dust  at 575 K and  120 K
with covering factors of $R=0.15$ and 0.006, respectively. 

WX CrA is one of the few stars in the sample with IRAS fluxes, especially at
12$\mu$m, that are much greater than Spitzer values. The IRAS fluxes require a
blackbody at 700 K with a $R=0.49$. This also accounts for the L and N
magnitudes but not the M magnitude from Kilkenny \& Whittet (1984). Glass's
(1978) M magnitude is 0.8 magnitudes fainter and falls close to the fit to
the IRAS fluxes.  Glass comments:  `Long-term variations, not closely associated
with visible-region behaviour, were observed at L.' By extension, we infer these
variations occur at longer wavelengths too.  

{\bf V3795 Sgr:} This star observed by Spitzer at maximum light is a warm 
`minority' RCB which has undergone  only two declines in the last  20 years with
each lasting about five years; Jurcsik (1996) gives the inter-fade period as
6000 days, one of the longest in her sample. 

VRI and JHK photometry (Table 2) was obtained almost simultaneously with the
Spitzer observations.  A B magnitude is estimated by combining the V from Table
2 with the (B-V) from Kilkenny et al. (1985). JHK magnitudes from Table 2 and
2MASS  are in fair agreement. There are slight differences between these values
and those provided by Feast et al. (1997) who also gave a N magnitude. Asplund
et al. 1997 estimates  E(B-V) = 0.79 but Feast et al. adopted E(B-V)= 0.45;
here, the higher value is assumed. 

The stellar blackbody temperature is set at 8000 K, the effective temperature
estimated from spectroscopy by  Asplund et al. (2000). A dust temperature of 610 K
and a covering factor $R=0.31$  fit the dereddened  Spitzer fluxes. The IRAS 12
$\mu$m and 25 $\mu$m fluxes, which are only slightly greater than Spitzer values,
require a dust blackbody of 720 K and a covering factor $R=0.54$.  This accounts
quite well  for the flux at L from Feast et al. (1997) obtained when the star was at
maximum. Feast et al.'s K magnitude is 0.5 magnitude brighter than the 2MASS  and
Table 2 values suggesting that warm dust affected their measurement at K.  

{\bf V1157 Sgr:} This cool RCB (Lloyd Evans et al. 1991) has undergone at least three
minima in the last 20 years. According to the more recent ASAS-3 database, V1157
Sgr has experienced at least five minima of more than 2 magnitudes in the last nine
years. When Spitzer observed the star it was about two magnitudes below maximum light
(Table 2).

Available photometry is limited to that in Table 2 and the 2MASS JHK results.
Adopting the latter as a measure of the star at maximum light and with an E(B-V)
of  0.3 estimated from a comparison of colors with those of HdC stars, a fit
suggests a stellar blackbody temperature  4200 K and dust blackbodies at 770 K
and 120 K with covering factors of $R=0.59$ and 0.007, respectively.  The 2MASS K
magnitude received approximately equal contributions from the star and the
dust. The IRAS 12$\mu$m and 25 $\mu$m fluxes, which are greater than the
Spitzer fluxes require dust at 850 K. 

{\bf Y Mus:} Y Mus, a warm RCB, has not experienced a decline in more than 20
years. Jurcsik (1996) gives the inter-fade period as 15300 days, the third
longest in her list. Feast et al. (1997) note a brief decline in 1953 reported
by Siedel (1957). 

Simultaneous ground-based photometry was not obtained but in light of the star's
insistence on remaining at maximum light, photometry in the literature may be
used to construct the SED. UBVRI photometry is taken from Kilkenny et al.
(1985). JHK 2MASS measurements agree with a single observation by Feast et al.
(1997) who also provide an L magnitude.  Kilkenny \& Whittet (1984) give M and N
from 1983, the IRAS epoch. An interstellar E(B-V) = 0.5 is adopted (Feast et al.
1997;  Asplund et al. 1997). 

Reddening corrected fluxes are well fit by a stellar blackbody at 7200 K and a
dust blackbody at 395 K with the low covering factor $R=0.01$ (Figure 7). 

Strikingly, the infrared excess from IRAS fluxes are considerably stronger: IRAS
12 $\mu$m and 25 $\mu$m fluxes are 4.5 and 2.9 times the Spitzer values,
respectively. The 1983 MN observations (Kilkenny \& Whittet 1984) span the IRAS
fit: M is about 50\% stronger and N is about 20\% weaker than the IRAS fit. L
(Feast et al. 1997) is matched by this fit to the IRAS fluxes. Only RT Nor and
UV Cas have comparable ratios of IRAS to Spitzer fluxes. The fit to the IRAS
fluxes and UBVRIJHKMN photometry gives a dust temperature of 590 K with a
covering factor $R=0.07$.

Evidently, this infrequently declining RCB had  an unusually weak circumstellar
dust  shell at the time of the Spitzer and IRAS observations.

{\bf V739 Sgr:}   This is a cool RCB discovered by Lloyd Evans et al. (1991).  The V
magnitude (Table 2) agrees very well with the value listed in the ASAS-3 database
indicating that the star was at maximum light at the time of the Spitzer 
observations. Sparse AAVSO measurements of the visual magnitude across 20 years suggest
the star may be a frequent decliner. This is corroborated by the ASAS-3 database,
which shows at least four declines in the last nine years. 

Photometry is available  at VRIJHK from Table 2. The interstellar reddening is
assumed to be E(B-V)=0.5, the estimate for VZ Sgr in the same direction.  A fit
to the dereddened photometry and Spitzer fluxes gives a stellar blackbody of
5400 K with the K-band dominated by the dust emission. Spitzer fluxes are well
fit with a 640 K blackbody with a covering factor $R=0.59$ and there is a hint of
a cooler blackbody at 100 K with $R\simeq 0.005$. The K flux which is  not
primarily from the star suggests the presence of dust hotter than 640 K. The
IRAS 12 $\mu$m flux is within a few per cent of the Spitzer flux. This close
correspondence and the fact that the 25 $\mu$m IRAS  flux is of low quality makes the
fit to the IRAS fluxes more uncertain. Two blackbodies of 900 K and 700 K with
$R=0.64$ and 0.228, respectively, can fit the K band flux and the IRAS photometry
simultaneously.

{\bf VZ Sgr:} This warm  `minority' RCB  has experienced several declines of
differing depths in the last 20 years; Jurcsik (1996) gives the inter-fade
period as 1300 days. At the time of the Spitzer observations, VZ Sgr was
recovering from a deep prolonged decline and still several magnitudes below
maximum.  

Stellar fluxes for the star at maximum light are taken from the literature:
UBVRI (Kilkenny et al. 1985), JHK (Feast et al. 1997; 2MASS), and L (Feast et
al. 1997). The  interstellar reddening is  E(B-V)=0.30 (Feast et al. 1997). 

The fit to the dereddened photometry and Spitzer fluxes gives a stellar
blackbody at the spectroscopic effective temperature of 7000 K (Asplund et al.
2000)  and dust at temperatures of 700 K and 140 K with covering factors of
$R=0.17$ and 0.008, respectively (Figure 8).  The IRAS 12 $\mu$m and 25 $\mu$m
straddle the Spitzer fluxes. Additionally, the L magnitude from June 1995 (Feast
et al. 1997) is well matched by the dust's contribution at 700 K with a small
contribution from the star. The K magnitude is slightly contaminated by dust
emission. These data suggest that the  dusty envelope has maintained a high
degree of uniformity over decades. 

{\bf U Aqr:} U Aqr, a cool RCB in the Galactic halo, is distinguished  by its
extraordinary enrichment of light $s$-process (e.g., Sr) nuclides but a
near-normal abundance of heavy $s$-process (e.g., Ba) nuclides  (Bond et al.
1979; Vanture et al. 1999).  Jurcsik (1996) puts the inter-fade period at 1850
days but in the last 20 years, U Aqr has spent about 10 years below maximum
light.  When the Spitzer observations were obtained, U Aqr was about 0.6
magnitudes in V below maximum light in a slow recovery from a deep
decline.\footnote{High-resolution spectra in sample regions of the K band 
obtained less than two months before the Spitzer observations show stellar
molecular absorption features and, therefore, the star not the dust was the
dominant contributor to the K band (Garc\'{\i}a-Hern\'{a}ndez et al. 2010).} 

The SED is constructed from photometry acquired at maximum light.  The
interstellar reddening for this halo star about 10 kpc above the Galactic plane
(Lawson \& Cottrell 1997) is slight: E(B-V)=0.05  (Rao 1995 in Asplund et al.
1997; Feast et al. 1997). Photometry is from the following sources: UBVRI
(Lawson et al. 1990; Marang et al. 1990), JHK (2MASS), JHKL (Feast et al.
1997). 

The blackbody combination of 5000 K for the star and 475 K and 140 K for the dust
fits the data  with covering factors of 0.23 and 0.021 for the cool blackbodies. 

The IRAS fluxes straddle the Spitzer spectrum: the IRAS flux at 12 $\mu$m
exceeds its Spitzer counterpart but the IRAS upper limit at 25 $\mu$m  is less
than the Spitzer value. A fit to the IRAS data suggests a dust blackbody at 560
K with a covering factor of 0.37. This warmer black body fits the L magnitude
from Feast et al. (1997).

{\bf MACHOJ181933:} This is a cool RCB discovered by Zaniewski et al. (2005).
Comparison of photometry in Table 2 and from the discovery paper show that the
star was at maximum light at the time of the Spitzer observations.  Nothing is
yet known about the frequency of declines. 

VRI (Table 2) and JHK (2MASS) photometry are available. The interstellar
reddening is uncertain but not negligible; the star is in the direction of the
Galactic Bulge. Zaniewski et al. (2005) put E(B-V) at 1.0. Here, we adopt
E(B-V)= 0.5.

A fit to VRIJHK fluxes and the Spitzer spectrum is obtained with a stellar
blackbody of 4200 K and blackbodies of 695 K and 140 K with covering factors  of
0.48 and 0.022, respectively, for the latter two blackbodies. The stellar
temperature is similar to that of S Aps. 

{\bf ES  Aql:} ES Aql, a cool RCB  (Clayton et al. 2002) declines quite frequently:
AAVSO and ASAS-3 observations show a major decline about every year. Not
surprisingly, Spitzer caught ES Aql  recovering from a deep decline; it  was at V=12.3
or about 0.8 magnitudes below maximum light.  There is no multicolor photometry for ES
Aql at maximum light. 

BVRI at maximum are inferred from Clayton et al.'s Table 1 and discussion. JHKL
magnitudes\footnote{Note that the 2MASS magnitudes of ES Aql are not saturated
as stated by Clayton et al. (2002) (G. Clayton 2011, private communication).}
are also from  Clayton et al. (2002).  Adopting an interstellar reddening
E(B-V)=0.32 (Clayton et al. 2002) and a stellar blackbody of 4500 K fitted to
the dereddened BVRI, the Spitzer fluxes are fit with a 700 K blackbody and a
covering factor of 0.49  (Figure 9). The L magnitude from 1997 June fits the SED
composed of the 4500 K and 700 K blackbodies which is expected because for
active stars like ES Aql the L magnitude is little affected as a star goes from
maximum to minimum. IRAS 12 $\mu$m and 25 $\mu$m fluxes are within 10\% of
their Spitzer values.

{\bf FH Sct:} This warm RCB  has been largely ignored by observers probably
because  it appears within (but beyond) the Galactic cluster NGC 6694 (M 26). At
the time of the Spitzer observations, FH Sct was at 13.00 (Table 2) in
agreement with the ASAS-3 database and suggesting that it was about 0.5
magnitudes below maximum.

VRI are taken from Table 2.  2MASS JHK are assumed to refer to maximum light. A
reddening E(B-V)=1.0 is adopted from a comparison of (B-V) colors of FH Sct 
with RCBs of similar temperature.

A stellar blackbody of 6250 K (the spectroscopic effective temperature - Asplund
et al. 2000) is adopted and the Spitzer spectrum is fitted by blackbodies of 540
K and 140 K with covering factors $R=0.10$ and $R=0.002$, respectively. FH Sct
is a rare case where the IRAS 12 $\mu$m flux is less than the Spitzer flux. The
IRAS 25 $\mu$m flux of moderate quality is similar to the Spitzer value. 

{\bf SU Tau:} This warm RCB frequently experiences declines;  SU Tau has been
three or more magnitudes below maximum light for nearly half of the last 20
years.  Jurcsik (1996) gives the inter-fade period as 1200 days.  Despite its
propensity to live below maximum light, the photometry reported in Table 2 shows
that it was at maximum light during the Spitzer observations. 

The SED was constructed from the following: UBV (Fernie et al. 1972),  VRI
(Table 2), and  JHKL (Table 2). The 2MASS JHK are 1.4, 1.0, and 0.4
magnitudes,  respectively, fainter than those in Table 2 but were measured when
SU Tau was recovering from a deep decline.  Interstellar reddening  of
E(B-V)=0.50 (Glass 1978; Feast et al. 1997) is adopted.

Adopting a stellar blackbody at 6500 K, the spectroscopic effective temperature
(Asplund et al. 2000), the Spitzer spectrum and the L flux are matched with a
635 K blackbody with a covering factor $R$=0.45. A fit to the UBVRIJHK fluxes
gives a stellar blackbody temperature of 6500 K, the spectroscopic effective
temperature  (Asplund et al. 2000). IRAS 12 $\mu$m and 25 $\mu$m fluxes are both
slightly higher than Spitzer fluxes. 

{\bf DY Per:} This RCB  advertised to be `the coolest metal-poor' RCB (Yakovina
et al. 2009) declines  at approximately two year intervals. At the time of the
Spitzer observations, DY Per was at V$\simeq$ 11.8 or  about 0.5 magnitudes
below maximum light and at the beginning of a decline that eventually reached a
depth of more than five magnitudes below maximum light. Yet it is not very
clear if DY Per-like stars really are RCB stars (e.g., Alcock et al. 2001).

We assemble published  VBRIJHKLM magnitudes made at or near maximum light 
(Za\v{c}s et al. 2007; Alksnis et al. 2009) and adopt the reddening E(B-V)=0.48
(Za\v{c}s et al. 2007).  The 2MASS JHK magnitudes are about 0.4 magnitudes
fainter than the adopted values, an indication that they were obtained during a
decline.\footnote{Za\v{c}s et al. report that DY Per has a visual companion, a
G0 dwarf at a separation of 2.5 arc sec. This is unlikely to contribute to the
SED.}  

A fit to the dereddened photometry and the Spitzer spectrum with a stellar
blackbody at 3000 K  (Yakovina et al. 2009) requires blackbody at 1400 K for the
dust. The IRAS fluxes are so similar to Spitzer values that a separate fit was
not made.  With this combination,  JHKL are dominated by the dust contribution. 
The stellar temperature is somewhat hotter than Yakovina et al.'s (2009)
extimate  of 2900-3100 K from model atmosphere predictions  fitted to a
4300-7300 \AA~SED but this apparent discrepancy may be due to the presence of 
strong molecular absorption bands which are not taken into account by a
blackbody representation of a SED. More significantly, Tenenbaum et al. (2005)
show that the first- and second-overtone CO bands are seen in the K and H bands,
respectively, showing that the star dominates the flux in these bands. The
M-band  flux  is well below that implied by the 1700 K blackbody and may be due
to strong CO fundamental band absorption. The IRAS and Spitzer fluxes are very
similar.

The Spitzer spectrum has a short wavelength cutoff at 10 $\mu$m and, therefore,
no information is provided about the 6-10 $\mu$m feature widely seen in RCB
spectra. However, DY Per's spectrum shows a sharp emission feature at 11.3
$\mu$m on a `pedestal' extending from about 10 $\mu$m to 13.5 $\mu$m.

{\bf V517 Oph:} Kilkenny et al. (1992) identified V517 Oph as  cool RCB which is `very
active in the RCB sense', i.e.,  frequently in decline. This is corroborated by more
recent ASAS-3 measurements, which show several minima in the last nine years. The optical
spectrum shown by Kilkenny et al. shows great similarity with S Aps.  At the time of the
Spitzer observations, the star at V=14.05 (Table 2) was about 2.4 magnitudes below maximum
light. 

Photometry in Table 2 enables the stellar SED to be estimated as it was at the time of the
Spitzer observations  but it is more valuable for understanding the relation between dust
and star to analyse the SED  at maximum light. Unfortunately because the star is faint and
frequently in decline, measurements at maximum are rare.  UBV photometry reported by
Kilkenny et al. was, as they noted, obtained below maximum light: their brightest V=12.4
may be about one magnitude fainter than maximum light values from the AAVSO and
ASAS-3 databases. 2MASS JHK magnitudes are 2.2 (J), 1.5(H), and 0.8 (K) magnitudes
brighter than those in Table 2.  Kilkenny et al. estimate E(B-V)= 0.5--0.6.

A fit to the 2MASS photometry is possible with a stellar blackbody of 4100 K
and a dust blackbody of 850 K with a covering factor of 0.84. This fit accounts
quite well for the L flux (Table 2) and a substantial amount of the flux at K is
from the dust. The IRAS fluxes are somewhat larger than Spitzer values: 20\% at
12 $\mu$m and 10\% at 25 $\mu$m over Spitzer values. 

{\bf V CrA:}  This  warm minority RCB  was observed by Spitzer at maximum light
following a deep decline two and a half years previously. The star frequently
declines; Jurcsik (1996) gives the inter-fade period as 900 days and only three
stars in her sample decline more often. 

Photometry at maximum light is taken from the literature: UBVRI (Lawson et al.
1990) and JHKL (Feast et al. 1997; Glass 1978). 2MASS JHK measurements are
fainter by 0.2 (J), 0.5 (H) and 0.7 (K) magnitudes than Feast et al.'s mean
values. Kilkenny \& Whittet (1984) measured the M and N magnitudes at a time
when the star was at least three magnitudes below its maximum visual
brightness.  V CrA is only slightly reddened: E(B-V) = 0.14  (Rao 1995 in
Asplund et al. 1997). 

A  fit to the  UBVRI fluxes and the Spitzer fluxes with a 6500 K stellar
blackbody, the spectroscopic effective temperature (Rao \& Lambert 2008), calls
for dust blackbodies at 550 K and 150 K with covering factors of 0.38 and
0.020,  respectively.  This fit does not account for the 2MASS HK and the above
LMN magnitudes, all of which are consistently brighter than the Spitzer
spectrum. Clearly, the infrared flux varies considerably.  The IRAS 25 $\mu$
flux and the KLMN fluxes, all observed at different times,  are  reasonably well
fit with a 6500 K blackbody and dust at 1600 K, 900 K, and 550 K with covering
factors of  $R= 0.23, 0.37,$ and 0.37, respectively. However, the IRAS 12 $\mu$m
flux is about $\sim$15\% larger than suggested by our blackbody fitting. This
suggests that the dust may be distributed in a disk rather than in a simple
discrete cloud or shell. Indeed, Bright et al. (2011) report possible
asymmetries in the V CrA's circumstellar dust shell from VLTI interferometric
data at 10 $\mu$m.

{\bf RZ Nor:} RZ Nor is a warm RCB notable for the presence of lithium.  Jurcsik (1996)
gives the inter-fade period as 1100 days, a typical value. According to the ASAS-3
database, the Spitzer observations were obtained  when RZ Nor was at maximum light.

Photometry at maximum light is provided by Glass (1978) for JHKL, Kilkenny et
al. 1985) for UBVRI and by Feast et al. (1997) for JHKL. Kilkenny \& Whittet
(1984)  observed at M and N when the star was about two magnitudes below
maximum.  Also, the 2MASS observations were obtained at a minimum when  J was
3.8 and K was 0.9 magnitudes fainter than Feast et al.'s estimates for maximum
light. A reddening of E(B-V)=0.5 is adopted (Glass 1978; Kilkenny \&
Whittet 1984; Rao 1995 -- see Asplund et al. 1997; Feast et al. 1997).

Fluxes from U to H are fit by a stellar blackbody of 5000 K, a temperature lower than
the spectroscopic effective temperature (Asplund et al. 2000). Spitzer fluxes are fit
with blackbodies of 700 K and 320 K with a hint of a 25 K blackbody introduced to
account for a flux increase longward of 34 $\mu$m.  The covering factors are 0.53 and
0.035 for the 700 K and 320 K blackbodies, respectively. IRAS fluxes for 12 $\mu$m
and 25 $\mu$m can be fit with a similar combination of blackbodies at 700 K and 300 K
and covering factors of 0.53 and 0.040, respectively.

{\bf RT Nor:} RT Nor,  a warm RCB, was at maximum light when observed by
Spitzer. Its last decline occurred about 15 years before. This fact
demonstrates the statistical nature of RCB declines because Jurcsik (1996) gives
the inter-fade period as short as 1950 days. 

Photometry of RT Nor at maximum light is assembled from the literature: UBVRI 
(Kilkenny et al. 1985), JHKLM Glass (1978), JHK (2MASS), JHKL (Feast \& Glass
1973),   MN (Kilkenny \& Whittet 1984). Although limited in coverage, the
photometry at LMN suggests strong variability. For example, Feast et al. (1973)
give L=6.42 whereas Glass (1978) reports a range from 7.6 to 8.3 from five
observations. Similarly, Kilkenny \& Whittet (1984) give M=5.8 but Glass
measured M=4.8.   Interstellar reddening of E(B-V)=0.39 (Glass 1978; Kilkenny \&
Whittet 1984; Rao 1995) is assumed.

A stellar blackbody of 6700 K and  dust blackbodies of 320 K and 130 K with 
remarkably low covering factors of 0.01 and 0.001, respectively, provide a
satisfactory fit to the UBVRIJK and Spitzer fluxes (Figure 10 - lefthand panel).
It is not surprising that the excess emission at 6--10 $\mu$m is also unusually
weak for RT Nor. 

Dust emission was much stronger at the time of the IRAS observations. The ratio
IRAS to Spitzer fluxes is a factor of five  at 12 $\mu$m and three at 25 $\mu$m.
A fit to the IRAS fluxes requires a blackbody of 500 K and a covering factor of
0.11 (Figure 10 -- righthand panel). This fit does not account for the LMN
fluxes: the LM fluxes are greater than the 500 K blackbody but at N the flux is
less than that blackbody. A fit solely to the UBVRIJHKLMN fluxes calls for the
stellar blackbody to be accompanied by a dust blackbody of 920 K with a
covering factor of 0.11.  The limited observational data suggests that RT Nor
experienced a increase in its infrared excess around 1980.

{\bf RS Tel:} RS Tel, a cool RCB, was observed by Spitzer shortly before it
underwent a prolonged decline. This decline apart, RS Tel has experienced no
major declines  and only one minor decline of about two and a half visual
magnitudes in the last two decades. However, Jurcsik (1996) gives the inter-fade
period as 1200 days.

Photometry at maximum light is taken from the literature:  UBVRI (Kilkenny et
al. 1985; JHKLM (Glass 1978),  Goldsmith et al. 1990), JHK (2MASS), JHKL (Feast
et al. 1997) and MN (Kilkenny \& Whittet 1984). The interstellar reddening is
taken to be E(B-V)= 0.17 (Feast et al. 1997; Kilkenny \& Whittet 1984; Rao 1995
-- see Asplund et al. 2000; Bergeat et al. 1999).

Optical and Spitzer fluxes are are well fit by a stellar blackbody of 6750 K, the
spectroscopic effective temperature, and blackbodies of 720 K and 130 K with
covering factor of 0.25 and 0.005, respectively (Figure 11).

The IRAS 12 $\mu$m and 25 $\mu$m fluxes are just 10\% greater than Spitzer
values, suggesting a slightly cooler dust blackbody of 620 K and a
covering factor of 0.22. Also, the LMN fluxes are quite well reproduced by the
combination of the 6750 K and 720 K blackbodies. Since the LMN and IRAS
photometry were obtained decades prior to the Spitzer spectrum, the ability of a
single fit to match all the observations implies long-term uniformity of the
circumstellar envelope.

{\bf V482 Cyg:} The RCB V482 Cyg has a K5III companion only 6 arc sec away
(Gaustad et al. 1988; Rao \& Lambert 1993).  The star was at maximum light when
observed by Spitzer. It has experienced just two declines in 20 years, one 14
and the other 19 years ago. Jurcsik (1996) gives the inter-fade period as 3400
days.

Photometry at maximum light is assembled from the literature: BVRI (Rosenbush
1995), JHK (2MASS), JHKL (Gaustad et al. 1988). An interstellar reddening of
E(B-V)=0.5 is adopted (Rao \& Lambert 1993). 

The BVRIJHK fluxes are well fit with a blackbody of 4800 K which with blackbodies
at 500 K and 100 K with covering factors of 0.03 and 0.001, respectively, combine
to fit the Spitzer fluxes.  The stellar blackbody is considerably cooler than
the 6750 K effective temperature provided by Asplund et al. (2000) from their
abundance analysis. 

The IRAS 12 $\mu$m and 25 $\mu$m fluxes are 80\% and 10\%, respectively, greater
than Spitzer fluxes. A fit to the BVRIJHK and IRAS fluxes with the 4800 K
stellar  blackbody demands a 650 K blackbody and a covering factor of 0.09. 
Neither the IRAS nor the Spitzer fit accounts for the flux at L measured in
1984. This requires a blackbody of about 1800 K, a temperature hotter than
the sublimation temperature of carbon soot, which may result from contamination
by the K5 companion.

{\bf MV Sgr:} This hot RCB star, according to De Marco et al. (2002), falls
outside the reach of `current stellar evolutionary models', i.e., the star is
not readily attributable to either the FF or DD scenarios. MV Sgr  rarely
experiences a decline; the last recorded decline was in the 1950s (Hoffleit
1959).  Thus not suprisingly, Spitzer observed the star at maximum light. 

Photometry is taken from the following sources: UBVRI (Kilkenny et al. 1985;
Goldsmith et al. 1990), JHKLMN (Kilkenny \& Whittet 1984)  and JHK (2MASS).
DeMarco et al. estimate E(B-V)=0.43.  

De-reddened BVRI fluxes are well fit with a blackbody of temperature 15400 K, the
effective temperature estimated from spectroscopy by Jeffery et al. (1988) but other
temperatures will fit these fluxes in the Rayleigh-Jeans tail.  The Spitzer spectrum
is matched with  blackbody at 205 K and a covering factor of 0.18.  The IRAS fluxes
are reproduced by a slightly hotter 235 K blackbody with a slightly larger covering
factor. This increase also fits the N flux but the M flux exceeds the fit by a factor
of about 50 per cent. 

The limited photometry at JHKLMN indicates the presence of a variable source
with a temperature of 1500K or so. For example, the JHKL fluxes require in
addition to the 15400 K and 205 K blackbodies a blackbody at 1500 K with a covering
factor of 0.33. A fit to IRAS and MN fluxes requires 1500 K and 235 K blackbodies
together with the stellar 15400 K blackbody to fit BVRI. This combination
predicts too strong a flux at K. This contribution at about 1500 K may come
from fresh production of carbon soot.

{\bf RY Sgr:} For RY Sgr, a warm RCB, deep declines occur about every 4 years
and the Spitzer observations were made when RY Sgr was about three magnitudes
below maximum light. Jurcsik (1996) gives the inter-fade period as 1400 days.

The star is represented by a 7200 K blackbody; the spectroscopic effective
temperature is 7250 K (Asplund et al. 2000). The Spitzer fluxes are fit with  a
blackbody of 675 K and a covering factor of 0.20. 

Fluxes measured by ISO from 4 $\mu$m to 26 $\mu$m exceed Spitzer values and
suggest a dust blackbody of 820 K with a covering factor of 0.38. IRAS fluxes
which  exceed both ISO and Spitzer values are fit with a dust blackbody of 870 K
and a covering factor of 0.76.  It is tempting in the sequence of declining
blackbody temperatures and covering factors from IRAS to ISO to Spitzer to see
an evolutionary sequence from IRAS in 1983 to  ISO in 1997  and then to Spitzer
in 2004 October. In light of the frequency of declines, an evolutionary
interpretation is probably too simplistic. Bright et al. (2011) report the
apparent presence of asymmetric circumstellar material around RY Sgr.

{\bf V854 Cen:} V854 Cen, a most active warm RCB,  has the shortest known
inter-fade period (370 days) of the RCBs studied by Jurcsik (1996).  Spitzer
observations were obtained when V854 Cen was at a visual magnitude between five
and six magnitudes below maximum.  An important distinguishing mark is its
modest H-deficiency among RCBs: hydrogen is deficient by a factor of only 100 to
1000 (Asplund et al. 1998). 

Published photometry for  maximum light  is available from Lawson et al. (1999)
for UBVRI  and Feast et al. (1997) for JHKL. Interstellar reddening E(B-V)=0.07
is adopted (Feast et al. 1997). 

BVRI are quite well fit with a blackbody of 6750 K, the spectroscopic  effective
temperature (Asplund et al. 1998). The Spitzer  spectrum is fit by a blackbody
combination of 900 K and 140 K with covering factors of 0.32 and 0.03,
respectively. This fit accounts fairly well for the HKL at minimum light, the
phase at which V854 Cen was when observed by Spitzer. 

IRAS fluxes are stronger than Spitzer values and similar to the ISO values
(Lambert et al. 2001). The fit to IRAS (and ISO) fluxes requires a 1100 K
blackbody with a covering factor of unity. The HKL fluxes at optical maximum are
dominated by dust emission; the flux increases with increasing wavelength. This
fit accounts quite well for the JHKL fluxes at maximum light. Indeed, the ISO
observation was made when V854 Cen was only slightly below maximum light.
Very recent high angular resolution interferometric observations of V854 Cen
show the possible presence of asymmetries in the circumstellar envelope (Bright
et al. 2011).

{\bf UW Cen:}  UW Cen, a warm RCB, has an inter-fade period of 1100 days
(Jurcsik  1996).  A reflection nebula illuminated by the star was discovered by
Polacco et al. (1991) and further studied by Clayton et al. (1999).  Spitzer
observations were obtained at maximum light during an  approximately 18 month
restoration to maximum light between an earlier deep decline lasting almost a
decade and the next decline  which ended early in 2010.

Photometry for maximum light is taken from the literature: UBVRI  (Kilkenny et
al. 1985; Goldsmith et al. 1990), JHKLM (Glass 1978), JHKL (Feast et al. 1997;
Goldsmith et al. 1990), MN (Kilkenny \& Whittet 1984; Goldsmith et al. 1990).  
Interstellar reddening E(B-V)=0.32 is assumed (Glass 1978; Kilkenny \& Whittet
1984; Rao 1995 -- see Asplund et al. 1997; Feast et al. 1997).

A fit  to the dereddened fluxes  with a 7500 K  stellar blackbody, the
spectrocopic effective temperature (Asplund et al. 2000), calls for  blackbodies
at 630 K, 120 K and 50 K with covering factors of 0.44, 0.013, and 0.05,
respectively. Dust emission extends to the K band; the K-flux is greater than
that at H.\footnote{The fact that dust dominates the spectrum in the K band is
confirmed by K-band spectra obtained in 2007 February when UW Cen was at maximum
light (Garc\'{\i}a-Hern\'{a}ndez et al. 2009).} The fit to the Spitzer fluxes
accounts well for the L and M fluxes observed about 30 years earlier.  The N
flux is less than the Spitzer value by about 20\%.  The IRAS 12 $\mu$m flux is in
good agreement with the Spitzer value but the 25 $\mu$m flux is about 30\% higher
than the Spitzer measurement.

{\bf DY Cen:} DY Cen, a hot RCB, has not experienced a decline in 20 years,
although the maximum V magnitude seems to have fallen by 0.7 magnitude in 45
years (DeMarco et al. 2002). Jurcsik (1996) gives the inter-fade period as 6400
days. Lambert \& Rao (1994), on the basis of the abundance analysis by  Jeffery
\& Heber (1993), considered DY Cen to be a minority RCB. DeMarco et al. note
that DY Cen is likely to be related to be the other (cooler) RCBs. The star is
unusually H-rich for a RCB: H/He = 0.1 by number (Jeffery \& Heber 1993).  

Photometry in the literature provides the following: UBV (Pollacco \& Hill
1991), UBVRI (Kilkenny et al. 1985),  JHK (2MASS),   N (Kilkenny \& Whittet
(1984).   Interstellar reddening E(B-V)=0.47 is adopted (Jeffery \& Heber 1993;
Rao et al. 1993). 

A stellar blackbody of 19500 K, the effective temperature obtained by Jeffery \&
Heber from their spectroscopic analysis, is fitted to the reddening-corrected
UBVRIJHK fluxes. Since these bandpasses are in the Rayleigh-Jeans tail of a hot
blackbody, other fits are possible.  The Spitzer fluxes show that the dust is
predominantly cool with a blackbody of 272 K and a covering factor of 0.09.  
Emission features from a mixture of PAHs and C$_{60}$ molecules are prominent
(Garc\'{\i}a-Hern\'{a}ndez et al.  2011). IRAS fluxes are within about  10\% of
the Spitzer values.

{\bf R CrB:} R CrB is the most studied warm RCB. It was at maximum
when the Spitzer spectrum was obtained. Jurcsik (1996) gives the
inter-fade period as 1100 days.

The UBVRIJ fluxes are fit with a 6750 K blackbody, the spectroscopic effective
temperature (Asplund et al. 2000). The  10-20 $\mu$m Spitzer spectrum  is fitted
with a 950 K blackbody with a covering factor of 0.30. This fit also accounts
well for the K and L fluxes where the shell makes the dominant contribution to
the latter and the star to the former.  The IRAS 12 $\mu$ and 25 $\mu$m fluxes
are higher than Spitzer values. The ISO spectrum from 1998 January in fair
agreement with the IRAS fluxes was fit with a combination of two blackbodies:
1390 K and 610 K (Lambert et al. 2001). Rao \& Nandy (1986) fitted the IRAS
observations with a black body at 680 K and Clayton et al. (1995) fitted an IRAS
LRS spectrum with a blackbody at 650 K.  R CrB is a variable (out of decline) in
the infrared. There are variations of about one magnitude at K and of two
magnitudes at L. Strecker (1975) and Feast et al. (1997) suggest that the L
variations follow a period of about 1260 days. Thus, a fit of infrared - optical
SED requires a variable infrared excess.

{\bf V348 Sgr:} V348 Sgr is `the second most active RCB star (after V854
Cen)' (De Marco et al. (2002); Jurcsik (1996) gives the inter-fade period as 560
days. V348 Sgr is a hot RCB with an effective temperature  about 22000K (Jeffery
1995). De Marco et al. (2002) suggest with some reservations that V348 Sgr is a
final-flash post-AGB star becoming once again the central star of a planetary
nebula. Visible nebulosity is present (Pollacco et al. 1990) supporting the
evolutionary interpretation.  Spitzer observations were obtained when V348 Sgr
was undergoing a minimum and are discussed in more detail in Clayton et al.
(2011).

Photometry at maximum light is assembled as follows: BVRI (Heck et al. 1985), 
JHKLM (Glass 1978), JHK (2MASS). Observed fluxes are corrected for an
interstellar reddening of E(B-V)= 0.45 (Pollacco et al. 1990). 

Adopting a stellar blackbody at 20000 K, the Spitzer spectrum is fit with a
combination of 707 K and 100 K blackbodies with covering factors of 0.63 and 0.035,
respectively.  This fit  accounts well for the JHKLM fluxes from the 1970s and
the 2MASS JHK from 1998. Also, the IRAS fluxes  only slightly exceed their
Spitzer  counterparts.  Rao \& Nandy's (1986) fit to the IRAS 12, 25, 60, and
100 $\mu$m fluxes required warm dust (600 K) and cold dust (60-100 K), a mixture
very similar to the present fit to  the Spitzer observations obtained nearly 30
years later.  This constancy of the infrared emission over decades is not
surprising given that V348 Sgr is secondmost active RCB.

{\bf HV 2671:} De Marco et al. (2002) note that HV 2671 in the LMC and V348 Sgr
have almost identical optical spectra. The Spitzer spectra are, however, very
different (Figure 5; see also Clayton et al. 2011 for more details). De
Marco et al. provide JHK magnitudes and a reddening estimate E(B-V)=0.15.
Soszy\'{n}ski et al. (2009) report V and I for maximum light. On the assumption
that the stellar  blackbody temperature is 20000 K, the same as for V348 Sgr,
the fit to the JHK fluxes and the Spitzer spectrum calls for blackbodies at 590
K and 150 K with covering factors of 0.36 and 0.268, respectively. There is a
hint for the presence of an additional and very cool 40 K dust component with a
negligible covering factor. The dust distribution around HV 2671 differs
substantially from that around V348 Sgr, a not surprising result. 

\section{Probing the cloudy circumstellar environment}

Early investigations of the infrared emission from RCBs searched for variability
of that emission. Among the earliest studies of infrared variability, Forrest et
al.'s (1972) observations of R CrB through a six-magnitude visual decline showed
that the infrared flux from 3.5 $\mu$m to 11 $\mu$m remained essentially
unaffected by the decline. This result suggested that the dust cloud causing the
decline represents a small addition to the  warm circumstellar material already
in place. 

Long-term monitoring of the infrared excess of RCBs potentially offer particular
insights into the structure and growth of clouds in the circumstellar
environment. In this regard, Feast and colleagues at the  South African
Astronomical Observatory (SAAO) provided a valuable set of observations. For
example, Feast et al. (1997) report JHKL photometry of 12 RCBs for timespans of
up to 23 years. At JHK, the star generally dominated the observed flux but at L
the infrared excess from dust is significant. Observations at L along with
other evidence led to the proposal that dust is ejected into the circumstellar
environment in the form of `random puffs' (Feast 1979, 1986, 1996, 1997;  Feast
et al. 1997).

It is in the context of the `puff' model that we explore below  what may be
deduced from the relation between the Spitzer and IRAS fluxes and the dependence
of the covering factor $R$ on the frequency of optical declines. These relations
suggest that dust is ejected in puffs, often off the line of sight, and very
likely in most cases the ejection occurs approximately isotropically off the
star (i.e., at random directions).

\subsection{Spitzer versus IRAS}

A comparison of the Spitzer spectra with the IRAS 12 $\mu$m and 25 $\mu$m fluxes
provides insight into the variability of the emission from circumstellar dust on
a time scale of about 25 years. This  is potentially valuable because 29 of our
RCBs are in the IRAS catalog. In Figure 12, we present histograms of the ratios
of the IRAS to Spitzer fluxes  at 12 $\mu$m (left panel) and 25 $\mu$m (right
panel). 

With the exception of five outliers, the ratios at 12$\mu$m are  less than two:
the mean value is $r_{12} = 1.41\pm$0.35 from 22 stars.  A correction has to be
applied because the IRAS fluxes refer to a broad-band and assigned an effective
wavelength of 25 $\mu$m. The effective wavelength assumes the energy distribution
has the form $f_\nu \propto \nu^{-1}$ or $f_\lambda \propto \lambda^{-1}$. Given
that the Spitzer spectra are close approximations to a blackbody spectrum, one
may estimate a color correction according to a recipe provided in the IRAS
catalog (Beichman et al. 1988). This color correction reduces the IRAS catalog
entries by about 16 per cent or the ratio $r_{12}$ is 1.18. 

Setting aside the same five outliers, the ratio $r_{25} = 1.27\pm0.32$ from 21
stars: R CrB was not observed by Spitzer at 25 $\mu$m and the ratios for VZ Sgr
and U Aqr are set at the value given by the upper limit to their IRAS 25 $\mu$m
fluxes.  After the color correction, this value of $r_{25}$ is not sensibly
different from unity. 

For stars with extended infrared emission, the IRAS fluxes will be
systematically larger than Spitzer values because of the larger IRAS aperture.
The IRAS aperture was approximately two minutes of arc which is  larger than the
Spitzer aperture for the SH observations (4.7$^" \times 11.3^"$) used at 12
$\mu$m and for the LH observations (11.1$^" \times 22.3^"$) used at 25 $\mu$m.
Perhaps, coincidentally the 25 $\mu$m Spitzer fluxes through the larger LH
aperture are closer to their IRAS counterparts than the 12 $\mu$m fluxes through
the SH aperture. As we have already mentioned in Section 2.2, all RCBs in
our sample are point-like sources for Spitzer. IRAS and Spitzer fluxes agree
very well when astronomical sources are point-like for Spitzer (e.g.,
Garc\'{\i}a-Hern\'{a}ndez et al. 2007; 2009). Indeed, R CrB itself which is
known to be extended shows no important flux differences between IRAS and
Spitzer (also V348 Sgr with a PN of 30", Clayton et al. 2011). Thus, the
presence of the five outliers is not related to a possible extended emission in
these sources\footnote{Note also that extended emission is not seen in the
available Spitzer/IRAC images at 8 $\mu$m.}. Finally, note that possible
extended diffuse background emission is usually due to much colder dust (see
e.g., Cox et al. 2011), which emits at wavelengths longer than 25 $\mu$m (see
e.g., the case of the hot RCBs V348 Sgr and HV 2671; Clayton et al. 2011).

The IRAS-Spitzer comparison has, thus, shown that the circumstellar dust
emission for the great majority of the RCBs  is unchanged over the last couple
of decades. This is not a surprising  result given the extensive L band
photometric measurements on a fair sample of RCBs conducted at the SAAO over
several years (Feast et al. 1997). Attention is necessarily drawn to the five
outliers. Before discussing this quintet, we note the surprising result that
these outliers all with high values of $r_{12}$ and $r_{25}$\footnote{The
$r_{25}$ for WX CrA is 1.8, a value which does not qualify it as an outlier at
25 $\mu$m.} can not be matched with a similar number of outliers with remarkably
low values of $r_{12}$ and $r_{25}$. Three of the five have $r_{12}$ of between
five and six. A comparable outlier on the low-$r$ side of the histogram would
have a value of less than 0.2, but the sole RCB with $r_{12}$ of less than unity
is FH Sct with $r_{12}$=0.7. With respect $r_{25}$, the asymmetry is not quite
so severe. The lowest $r_{25}$ values are upper limits of 0.9 and 0.8 for VZ Sgr
and U Aqr, respectively to be compared with the highest two values of 3.3 and
2.9 for Y Mus and UV Cas, respectively. Is this presence of outliers on the high
$r$ and the paucity of outliers on the low $r$ side just a statistical fluke or
a hint at a long-term evolution in RCB dust shells? The latter seems unlikely
given that the evolutionary timescale is likely to be a few thousand years and
the interval between IRAS and Spitzer observations is less than 30 years. 

With respect to the outliers, a possible explanation is that these are RCBs
which rarely eject dust. Dust ejection rates may or may not be closely related
to the frequency of declines, i.e., formation and presence of dust along the
line of sight. Dust formation may or may not occur in preferential directions
such as an equatorial plane or as polar plumes. Next, we discuss the five
outliers in order of decreasing $r_{12}$: three -- Y Mus, UV Cas, and RT Nor --
have $r_{12} \simeq 5$ and two -- WX CrA and  RY Sgr  -- have $r_{12} \simeq
3$. 

{\bf Y Mus:} Y Mus has not experienced a decline in the nearly thirty years
covered by AAVSO records which begin about  1982 January.  The  covering factor
$R=0.009$ from the Spitzer observations is the lowest among our RCB sample. Even
the IRAS fluxes correspond to a low $R=0.07$. Thus, we conclude that Y Mus is
simply a poor producer of dust, i.e., the preferred axis for dust production is
not orthogonal to the line of sight.  One supposes that Y Mus is slightly more
active with respect to dust production than XX Cam which has not been observed
in decline at all and has no infrared excess out to 10 $\mu$m (Rao et al. 1980).

{\bf UV Cas:} At the time of the IRAS observations, UV Cas was considered to be
a typically dusty RCB with a covering factor $R=0.28$ but by the time that
Spitzer observed it the infrared fluxes had declined sharply and the covering
factor had dropped to $R=0.03$. The dust cloud responsible for infrared emission
at the time of IRAS did not cause a deep optical decline of which none have been
seen for 60 years.    Bogdanov et al.'s (2010) survey of JHKLM for 25 years from
1984 show a major  IR excess present and weakening at the time of the IRAS
observations with two weaker episodes in subsequent years.    Bogdanov et al.'s
KLM magnitudes from 1984 are acceptably fit by the blackbody fit to the
reddening-corrected IRAS fluxes (Figure 6).

{\bf RT Nor:} RT Nor closely resembles Y Mus in terms of its low covering
factors at the time of the  IRAS and Spitzer observations. RT Nor has
experienced (according to AAVSO records) only three or four optical declines of
three or more magnitudes in the last 50 years. None of the declines happened a
few years prior to the IRAS or Spitzer observations.  As in the case of Y Mus,
we suppose that RT Nor is simply  an infrequent producer of dust.  

{\bf WX CrA:} In contrast to Y Mus, UV Cas, and RT Nor, WX CrA is frequently in
decline; for example, in the thirteen years from late-1992 to late-2005 it was
almost always below maximum light, often by two or more magnitudes. The Spitzer
observation acquired at maximum light was preceded by about three years at
maximum light. The IRAS observations were similarly acquired at maximum light
and preceded by about a decade without optical declines according to AAVSO
records and Feast et al. (1997). Thus, the greater IRAS fluxes and higher
covering factor ($R=0.49$ versus $R=0.14$) relative to Spitzer fluxes must be
due to ejections of dust off the line of sight. Then, the fact that both $R$
values are fairly representative values suggests that ejection of puffs occurs
about as frequently as optical declines, i.e., there is no  strong  directional
dependence for  ejection of puffs.

{\bf RY Sgr:}  Infrared emission from RY Sgr provides higher covering factors
than found for WX CrA suggesting a higher dust ejection rate. At the time of the
Spitzer observations, RY Sgr was in decline but the covering factor $R$ was only
0.20. An above average $R$ (=0.76) was found from the IRAS observations. Major
optical declines occur at a frequency of about one every five years. A
distinguishing mark of RY Sgr is its periodic Cepheid-like variation in optical
light and the consequent variation in the infrared emission by the dust which is
heated  by optical light (Feast et al. 1997; Feast 1979, 1986).   The amplitude
of the effect at L is about 0.8 magnitudes or a factor of two. Without a
correction for this pulsational variation, comparison of IRAS and Spitzer fluxes
are not immediately interpretable solely in terms of dust ejection episodes.
Neverthless,  we suggest that RY Sgr behaves similarly to WX CrA in terms of
dust emission.  

\subsection{Covering factors and frequency of declines}

There is a not unexpected correlation between the covering factor $R$ and the
frequency of declines.  As a measure of the latter, we use the inter-fade period
$\Delta$T (in days) determined by Jurcsik (1996). In Figure 13, we show
$\Delta$T versus $R$. Apart from  the hot RCB MV Sgr, $R$ appears independent of
$\Delta$T for values of $\Delta$T $>$ 2000 days but increases steeply with
decreasing $\Delta$T below 2000 days. The mean $\Delta$T for stars with $R <
0.30$ is 5600 days from 13 stars with the subset of 5 stars with $R < 0.10$
giving a mean $\Delta$T = 10300 days.  For $R < 0.30$, the mean is $\Delta$T =
1000 days from 7 stars. 

The simplest interpretation of this correlation is that the inter-fade period is
roughly the time between ejection of puffs from all or most parts of the star
and that these puffs take on the order of 2000 days to move out to distances
where the dust temperature is lower than about 100 K. Ten stars in Table
3 were not studied by Jurcsik. Of these, the data on V magnitudes in the
last 9 years for six stars (V1783 Sgr, V1157 Sgr, V739 Sgr, ES Aql, FH Sct, and
V517 Oph) are available in the ASAS-3 database and appear to be consistent with
the correlation suggested by Figure 13. With the exception of V1783 Sgr (with
$\Delta$T$\sim$1000 days), the other five stars are frequently in decline (at
least 5 declines in the last 9 years). The data for the other four stars (Z Umi,
MACHOJ181933, DY Per, and HV 2671) are too sparse to estimate their $\Delta$T.

\subsection{On the back of an envelope}

As a guide to aspects of the infrared emission
by dust in the circumstellar shell, the following back-of-an-envelope
calculations are offered. 

The equilibrium temperature of a grey dust grain ($T_d(r)$) in an
optically thin circumstellar environment is given
by 

\begin{equation}
T_d(r) = (\frac {R_*}{2r})^{0.5}T_*
\end{equation}

where $R_*$ is the stellar radius, $r$ is the radial distance from
the stellar center, and $T_*$ is the stellar blackbody
temperature (Kwok 2007, p.314, eqn. 10.32).  For $T_* = 6000$ K, a
representative temperature for a warm RCB, the dust temperature
is 1320 K at 10 stellar radii and falls to 500 K at 50 stellar
radii. This temperature at 10 stellar radii is close to the
condensation temperature of carbon soot. The temperature at 50
stellar radii is fairly typical of the warm blackbody temperature
from the fit to the Spitzer spectra. 

If the velocity of the dust is expressed in units of 10 km s$^{-1}$ and
the radius of the star in units of 100 solar radii, the time for
dust to travel out from a distance $r_i/R_*$ to $r_f/R_*$ is
given by 

\begin{equation}
t = 0.22 \frac{R_{100}}{v_{10}}[\frac{r_f}{R_*} - \frac{r_i}{R_*}]
\end{equation}

where $t$ is expressed in years. For example, dust moving at 10 km s$^{-1}$ will
take about 10 years to move from 10 to 50 stellar radii assuming the RCB radius is
about 100 solar radii which is representative of yellow supergiants like the RCBs.
A velocity of 10 km s$^{-1}$ is a typical value for circumstellar gas around
normal dusty AGB stars. Very blue-shifted components to the Na D lines are seen at
the time of recovery from a deep minimum - say, 300 km s$^{-1}$ (Rao et al. 1999).
The evolution of the RCB spectra and lightcurves during declines are
consistent with dust that forms close to the stellar atmosphere and then is
accelerated to hundreds of km s$^{-1}$ by radiation pressure (e.g., Clayton et al.
1992; Whitney et al. 1993). There is also some evidence that the strength of the
blue shifted absorption seen in He I 10830 $\AA$ is inversely correlated with time
since the last decline (Geballe et al. 2009; Clayton et al. 2011, in preparation).
At present, the location of this gas and its relation to the thinning puff are
unknown. However, our results indicate that the dust detected at the Spitzer and
IRAS wavelengths is expanding at may be tens of km/s but not hundreds (see
below). In deep declines, the photospheric spectrum is often `washed out', an
effect attributed to scattering of photospheric light off the puffs in the
circumstellar environment. There is no direct way to measure the velocity of the
dust because of the lack of sharp spectroscopic features that provide radial
velocities. Indeed, the nearest method to measure the velocity of the dust is to
study the dust scattered photospheric spectrum of the star by moving (expanding)
dust cloud. Such a model has been shown by Herbig (1969), Kwok (1976), and in a
more detailed way by Van Blerkom \& Van Blerkom (1978), where they show the change
in profiles of stellar absorption lines by expanding dust and the redshifts
expected. We have measured such redshifts of the absoprtion lines in three minima
of R CrB including the current one. Our KECK high-resolution (30,000), high S/N
spectra obtained when the star is at V$\sim$15th show the scattered stellar
absorption spectrum very clearly and redshifts of about 25 kms$^{-1}$ relative to
the mean radial velocity of the star at maximum light, suggesting dust velocities
of this order. We have also broadened the normal light maximum spectrum and
matched to the scattered spectrum with such velocities, suggesting also that dust
is expanding at maybe tens of kms$^{-1}$ but not hundreds (Rao \& Lambert, in
preparation). Thus, if the puffs are moving radially away from the star, an
expansion velocity of 10-20 km s$^{-1}$ seems to be well justified. Bogdanov et
al. (2010) - by using calculations where the momentum couple the gas and dust in a
self-consistent procedure - estimate the characteristics of the stellar wind due
to radiation pressure on the dust in two RCBs. Consistently, with our suggested
expansion velocity, they compute gas and dust expansion rates of 8.8 and 15.6 km
s$^{-1}$ for the RCBs UV Cas and SU Tau, respectively, supporting our independent
estimation of the dust expansion velocity in RCBs and suggesting that the
dust detected by Spitzer and IRAS may not be close to the star where dust
formation seems to occur. Finally, note that with dust cloud complex motions of
10-20 km s$^{-1}$ and assuming a distance of 2 kpc, the 18' diameter cool (30 K,
at wavelengths beyond 60 $\mu$m) dust shell seen around R CrB (Rao \& Nandy 1986;
Guillet et al. 1986) would have left the star 256000 years ago. Thus, this cool
dust component may be the remmant of the stage when R CrB was a red giant for the
first time (Rao \& Nandy 1986).

Thus, a RCB experiencing  infrequent ejections of dust would be expected to show
significant variations in infrared flux on timescales of a decade or two,  i.e.,
differences should be seen in comparing IRAS and Spitzer fluxes for those stars
which rarely emit puffs. Conversely, stars ejecting puffs at a rate much shorter
than this timescale should show a quasi-constant infrared excess   where the
covering factor will depend in part on the total solid angle subtended by puffs
which in turn will depend on the number of ejection sites close to the stellar
surface and the angular expansion of a puff as it moves away from the star.  

These order of magnitude estimates for dust temperature and timescale
may be tested using the RCBs which are extreme outliers in the
histogram of the IRAS to Spitzer 12 $\mu$m flux ratios. These are stars
for  which one might expect a single puff to be present at a given
time and, hence the same puff may have been observed by IRAS and Spitzer.
  Two tests
are offered. 

First, equations (1) and (2) may be combined to express the timescale $t$
in terms of the blackbody temperatures from the IRAS and the Spitzer
fluxes. Assuming $r_{100}/v_{10} = 1$ and taking the temperatures from
Table 3, we predict a timescale of 13, 20, and 28 years for UV Cas,
Y Mus, and RT Nor, respectively, the three stars with the most extreme
decrease in 12 $\mu$m flux from IRAS to Spitzer. These estimates are
very similar to the time interval of about 25 years between the IRAS and
Spitzer observations. For the other two outliers, both with less
extreme ratios of the ratio of IRAS to Spitzer fluxes, the estimated
timescale is much less than the 25 year time interval but this
may be due to their higher frequency of puff ejection and, hence, the
presence in the circumstellar shell of more than a single puff. 

The second test uses the derived covering factors and the change in
these factors between the IRAS and Spitzer observations. On the assumption
that the physical size of the puff is not changing, the covering factor $R$
and dust temperature $T_d$ are related as $R \propto T_d4$. For the three
most extreme outliers in the IRAS to Spitzer flux ratio distribution, the
ratio of $R$ factors and the ratio of $T_d4$ are in fair
accord: the $R$-ratio and $T_d4$-ratios are  11 and 6, 8 and 5, and 8  and 6
for UV Cas, Y Mus, and RT Nor, respectively. Again, the indication is that
a single puff present at the time of the IRAS observations remained
unaccompanied at the time of the Spitzer observations but had been
driven to a larger radial distance. 

This test may be applied to the entire sample including outliers. In Figure 14,
the ratio of the Spitzer to IRAS blackbody temperatures is plotted versus the
ratio of the covering factors.  Given their characteristic optical variability,
it is remarkable that the RCB stars form a rather well defined trend in this
figure. On the assumption, as above, that the physical size of a puff does not
change as it moves away from its star, the $R \propto T_d4$ may be applied to
the sample and provides the evolutionary track  shown on Figure 14.  Stars near
the point corresponding to equal ratios in $T_d$ and in $R$ are identified as
stars whose circumstellar environment was essentially unchanged between the IRAS
and Spitzer observations. The extremity of the trend is set by the three extreme
outliers (UV Cas, Y Mus, and RT Nor) for which we have suggested the principal
puff present at the time of the IRAS observations evolved away from the star to
greater distances and, therefore, a lower covering factor and a lower
temperature. An interpretation of the fact that many stars connect the
normalization point and the location of the extreme outliers is that for these
stars too the principal puffs present for IRAS remained the principal puffs for
Spitzer but then at a greater distance, Of course, this interpretation has the
novel, if uncomfortable conclusion, that the majority of  our collection of RCBs
is now  experiencing a decline in their propensity to eject puffs of soot. Since
the trend is anchored by three extreme outliers, it is challenging to identify
systematic errors in either the IRAS or Spitzer data that might account for the
uncomfortable conclusion. 
 
A key question about the formation and ejection of the dusty puffs
is whether there is a preferred location for their formation with
respect to the star. (Their ejection path is presumed to be radially
outward above the point of formation.) We consider a few limiting
cases. 

{\bf There is a preferred location for puff formation:}  One imagines, for
example, such possibilities as active regions at one or both rotational poles,
along a latitudinal belt, or a long-lived active region (e.g., Wdowiak
1975; Soker \& Clayton 1999). The covering factor will likely be small for
ejection from a single active region but larger for fairly uniformly active
latitudinal belt.

In this circumstance, optical declines will be less frequent than the
time between ejection of puffs unless the preferred direction
intersects the line of sight to the star. In the extreme case, the RCB
will show an infrared excess but very rarely or never an optical
decline. 

If time between fresh puffs is long, the infrared excess will
move from near- to far-infrared wavelengths before ejection of
a fresh puff, i.e., there will be large variations in flux at L
and at Spitzer/IRAS wavelengths. 

At the other extreme when puffs are ejected frequently, the star will
show smaller variations in infrared excess with a broader distribution in
infrared wavelengths. 

The outliers such as UV Cas, Y Mus, and RT Nor are candidates for puff ejection
occurring -- currently -- from an active region which ejects puffs off the line of
sight. This region may or may not represent a preferred location in the long-term.
Rao \& Raveendran (1993) have suggested a preferred plane for dust arounf V854 Cen
from polarimetry during two deep minima. Clayton et al. (1997) from their
spectro-polarimetric observations of R CrB in a deep minimum suggested a dust disk
or torus that obscure the star with diffuse dust above the poles. On the other
hand, resolved dust shells around RCBs at visible and infrared wavelengths suggest
diferent shapes ranging from spherical symmetry to slightly elliptical (Guillet et
al. 1986; Walker 1986; Clayton et al. 1999; Bright et al. 2011). Thus prefered
locations are a possibilty in some cases and under some circumstances.

{\bf Puff ejection occurs from regions distributed isotropically over the
stellar surface:} If the number of active regions is small and ejection of
puffs infrequent, there will be large variations in the infrared excess and
long intervals between optical declines with possibly a low covering
factor. This state of affairs is generally equivalent to that expected from
a preferred location for puff ejection lying off the line of sight. Again,
the outliers such as UV Cas and friends fall in this category.

At the other extreme, the active regions may be many  and puff ejection
very frequent, the infrared variations will be small, the optical
decline rate will be high and the covering factor high. These circumstances
describe well a majority of the RCBs studied here.

These conclusions drawn from Spitzer and IRAS flux similarities and
differences echo those given earlier by Feast et al. (1997) from
their two decades of JHKL photometry of about a dozen RCBs. 
The J magnitude is set by the stellar flux and is sensitive to
optical declines. The L magnitude monitors the dust emission,
especially the warm and presumably freshly-formed dust; Feast (1997)
suggests there is no evidence for dust warmer than about 1500 K.
Not surprisingly, the L magnitude of a given RCB is variable with
an amplitude of up to three magnitudes with larger L variations
associated with the longer timescales. There is a very approximate
tendency for the larger L variations to come from stars with the longer
inter-fade periods,   a correlation consistent with our observation that the
larger covering factors are paired with the shorter inter-fade periods.
Significantly, Feast et al. (1997) conclude that `models involving
fixed geometry for the ejection of dust from the star appear to be ruled out
and the data support the random dust-puff model.' This remark derived from
the dust as measured by the L magnitudes refers pretty directly to
the formation of fresh dust quite close to a star.

\section{Dust around RCBs in different environments}

How different or similar is the  dusty circumstellar shell around RCB stars in
different environments (e.g., at different metallicities) such as the solar
neighbourhood and the Large Magellanic Cloud?

The properties of dust around RCBs in the Large Magellanic Cloud (LMC) have
been explored by Tisserand et al. (2009) using the photometry from Spitzer IRAC
and MIPs bands.  In various diagnostic diagrams such as the [24.0] absolute
magnitude versus  the [8.0]$-$[24.0] color (as estimated from Spitzer mid-IR bands
and the LMC's distance modulus), the LMC cool (i.e, DY Per-like),  warm and hot
RCBs are distinguishable.  The DY Per-like dust shells have the bluest colors
and  lowest absolute magnitudes with the hotter stars trending to higher
absolute magnitudes and redder colors.

For the RCBs in our sample, we obtained monochromatic fluxes at 8.0 and 24.0
$\mu$m from the reddening-corrected Spitzer spectra. We obtained magnitudes in
the system of IRAC [8.0] and MIPs [24.0] $\mu$m by using the appropriate flux
calibrations (Engelbracht et al. 2007) and IRAC manuals. It is to be noted that
our [8.0] and [24.0] magnitudes are not band averaged as is the standard
photometry used by Tisserand et al. (2009). However, systematic effects between
our magnitudes and  those from Tisserand et al. (2009) are thought to be small.
For the few stars where our Spitzer spectrum does not extend to 8 $\mu$m, we
used the blackbody fits that characterize the observed spectrum (Section 3) to
estimate the flux at 8 $\mu$m. In addition, for stars like V854 Cen and DY Cen
where the 6$-$10 $\mu$m emission features dominate the spectra, we used the
emission free continuum flux at the appropriate wavelengths.

For our comparison with the LMC's RCBs, we must estimate the distances to the
local RCBs. We have assumed M$_{bol}$=$-$5.0 $\sim$M$_{V}$ for most of the
stars, with the exception of the cooler RCBs (see below). Individual
distances are estimated from the reddening-corrected V magnitudes from which
M$_{bol}$ at 8 and 24 $\mu$m were obtained. For DY Per, the distance of 2.7 kpc
as estimated by Za\v{c}s et al (2007) has been adopted. Tisserand et al.
(2009) show that the M$_{V}$ of RCBs in the LMC change from -5 to -3.4 as a
function of the V-I color, particularly for V-I between 0.6 and 1.5 (their
Figure 3), suggesting that cooler RCBs have lower M$_{V}$s. We have used the V-I
color of the cooler RCBs in our sample to obtain M$_{V}$ estimates except for U
Aqr and MACHOJ181933. These estimates are used in determining the distances to
the individual stars and these distances are in turn used in estimating the
absolute magnitude at 24 microns. U Aqr is a halo star and the  M$_{V}$ of -5
seems to be justified (Cottrell \& Lawson 1998). MACHOJ181933 displays the
largest V-I color in our sample and we have assumed M$_{V}$=$-$5.0 for this
star. This is because Tisserand et al. (2009) show that there is not a unique
relation between M$_{V}$ and V-I color for V-I colors greater than 1.5. 

Figure 15 shows our sample of Galactic RCBs in the [24.0]-absolute magnitude
versus the [8.0]$-$[24.0] color index plane, along with the LMC RCBs  studied by
Tisserand et al. (2009). The latter data were taken from Table 6 of Tisserand et
al. (2009), in which a distance modulus of 18.5 was used for the LMC objects.

The only DY Per-like object in our sample is DY Per itself which merges  with
the the LMC DY Per-like objects, suggesting that the distance estimate of DY Per
is not too much off.  The other LMC RCBs and Galactic RCBs merge in this plot
over the  total range, suggesting that the dusty circumstellar shells  are of
similar nature. At the highest [24.0]$\mu$m luminosities are found the  hot RCBs
and the Galactic center cool RCB MACHOJ181933 at M([24.0]) = -14.37. However,
if we assume a distance modulus of 14.4 for the Galactic Center, then the
[24.0]$\mu$m luminosity is reduced to $-$11.06 (see Figure 15), which is
consistent with the other RCBs in our sample and implying that M$_{V}$ is not
$-$5.0 for MACHOJ181933. Tisserand et al. (2009) found that M$_{V}$ of RCBs
extend from $-$5.2 to $-$3.4 or even $-$2.6.  The RCBs UV Cas, Y Mus, and RT Nor
along with W Men (denoted as W in Figure 15) lie far below the [24.0]$\mu$m
magnitude of other RCBs although they have similar [8.0]$-$[24.0] colors. This
suggests that these stars have lower than average dust content. Indeed, UV CAs,
Y Mus, and RT Nor are just the three outliers (showing the highest IRAS/Spitzer
flux ratios; see Section 4) displaying the lowest (0.01$-$0.035) covering
factors in our sample, being very poor producers of dust.  In short, we conclude
that RCBs  in the LMC and the Galaxy have similar dusty circumstellar shells. 

\section{Conclusions}

Our almost-complete sample of Spitzer/IRS spectra of R Coronae Borealis
stars has been combined with multi-color photometry of a RCB at
maximum light to provide a spectral energy distribution from,
in general, the U or V band to 37 $\mu$m. Each SED has been
fit with a blackbody to represent the stellar flux and one, two, or
three blackbodies to represent the emission from the circumstellar
dust. A typical RCB emits about 30\% of the stellar flux in the
infrared. Although not discussed in this paper, emission features
superimposed on the combination of dust blackbodies are
with but a couple of exceptions limited to emission from 6-10 $\mu$m.
The exceptions discussed by Garc\'{\i}a-Hern\'{a}ndez et al. (2011)
are DY Cen and V854 Cen where emission is from PAHs and  C$_{60}$
molecules.

For the majority of the RCBs, there is fair agreement between the
Spitzer spectrum and the 12 $\mu$m and 25 $\mu$m fluxes from IRAS
from about three decades previously. There are five exceptions where the
IRAS fluxes are between three to five times those recorded by
Spitzer. Oddly, these outliers do not have counterparts for which the
IRAS fluxes are lower than the Spitzer values. 

Our results are consistent with the proposal that clouds of carbon soot form in
puffs above the surface of a RCB. There is evidence that puffs are formed
randomly and without a prefered direction. For some stars and especially for the
above five outliers a single puff may dominate the infrared emission. Our sample
of Galactic RCBs and those in the LMC share the same properties of their dusty
circumstellar shells, as evidenced by their total 24 $\mu$m luminosity and
[8.0]$-$[24.0] color index.

Although an important next step is to attempt modeling the radiative
transfer in the dusty circumstellar environment for our sample of
RCB stars, the value of long-term monitoring through infrared
photometry and spectroscopy should not be underestimated. 

\acknowledgments

We would like to thank the referee G. C. Clayton for suggestions that help
to improve the paper. We thank Michael Feast, Patricia Whitelock and Fred van
Wyk at the SAAO for observing stars at the time of our Spitzer observations. We
thank too N.M. Ashok of PRL, Ahmedabad for observing several RCB stars at the
Gurushikhar Observatory. We thank John Lacy and Amanda Bayless for  discussions
about modeling the dusty environment of RCB stars. D.A.G.H. acknowledges support
for this work provided by the Spanish Ministry of Science and Innovation
(MICINN) under the 2008 Juan de la Cierva Programme and under grant
AYA-2007-64748. N.K.R. would like to thank A. V. Raveendran for his help.
D.L.L. acknowledges support for this work provided by NASA through an award for
program GO \#50212 issued by JPL/Caltech and the Robert A. Welch Foundation of
Houston, Texas through grant F-634. Extensive use has been made of the AAVSO's
database of observations of RCB stars; we are most grateful to the many
observers who have and continue to contribute to this record  of the stars'
variability.  This work is based on observations made with the Spitzer Space
Telescope, which is operated by the Jet Propulsion Laboratory, California
Institute of Technology, under NASA contract 1407. Part of this work is based on
observations made with the IAC-80 telescope under the Spanish Instituto de
Astrof\'\i sica de Canarias Service Time. The IAC-80 is operated by the
Instituto de Astrof\'\i sica de Canarias in the Observatorio del Teide. 

{\it Facilities:} \facility{Spitzer: IRS; IAC80: Camelot; SAAO: }.

\clearpage

\begin{deluxetable}{lccccccccc}
\tabletypesize{\scriptsize}
\tablecaption{The RCB stars sample$^{a}$ \label{tbl-1}}
\tablewidth{0pt}
\tablehead{
\colhead{RCB star} & \colhead{RA$_{\rm J2000}$} & \colhead{DEC$_{\rm J2000}$} &
\colhead{Category} & \colhead{F$_{12}$} & \colhead{F$_{25}$} & \colhead{Obs.
Date} & \colhead{Var.$^{b}$} &
\colhead{Modules} & \colhead{Program} \\
\colhead{} &\colhead{} &\colhead{} &\colhead{} &\colhead{[Jy]} &\colhead{[Jy]}
&\colhead{yyyy/mm/dd}& \colhead{} & \colhead{} & \colhead{\#}
}
\startdata
UV Cas       &23:02:14.62 &$+$59:36:36.6 &A &  3.81     & 1.28      &2008/08/12& max     &SL,SH,LH& 50212\\
S Aps        &15:09:24.53 &$-$72:03:45.1 &C &  2.71     & 1.02      &2008/04/25& max     &SL,SH,LH& 50212\\
SV Sge       &19:08:11.76 &$+$17:37:41.2 &C &  3.29     & 1.66      &2008/05/26& max     &SL,SH,LH& 50212\\
Z UMi        &15:02:01.33 &$+$83:03:48.6 &C &  2.12     & 0.82      &2008/10/20& min[2]  &SL,SH,LH& 50212\\
V1783 Sgr    &18:04:49.74 &$-$32:43:13.4 &C &  3.20     & 1.26      &2008/04/25& max     &SL,SH,LH& 50212\\
WX CrA       &18:08:50.48 &$-$37:19:43.2 &C &  2.31     & 0.77      &2008/10/10& max     &SL,SH,LH& 50212\\
V3795 Sgr    &18:13:23.58 &$-$25:46:40.8 &AB&  4.17     & 1.80      &2008/04/25& max     &SL,SH,LH& 50212\\
V1157 Sgr    &19:10:11.83 &$-$20:29:42.1 &C &  3.21     & 1.17      &2008/06/02& min[2]  &SL,SH,LH& 50212\\
Y Mus        &13:05:48.19 &$-$65:30:46.7 &A &  1.02     & 0.35      &2008/04/25& max     &SL,SH,LH& 50212\\
V739 Sgr     &18:13:10.54 &$-$30:16:14.7 &C &  1.27     & 0.36      &2008/04/25& max     &SL,SH,LH& 50212\\
VZ Sgr       &18:15:08.58 &$-$29:42:29.4 &AB&  1.11     & 0.59      &2008/04/25& min[4]  &SL,SH,LH& 50212\\
U Aqr        &22:03:19.70 &$-$16:37:35.2 &C &  1.11     & 0.51      &2008/06/29& max     &SL,SH,LH& 50212\\
MACHOJ181933 &18:19:33.75 &$-$28:35:58.0 &C &  $\dots$  & $\dots$      &2008/04/25& max     &SL,SH,LH& 50212\\
ES Aql       &19:32:21.61 &$-$00:11:31.0 &C &  1.43     & 0.52      &2008/05/27& max     &SL,SH,LH& 50212\\
FH Sct       &18:45:14.84 &$-$09:25:36.1 &A &  0.62     & 0.49      &2008/06/02& max     &SL,SH,LH& 50212\\
SU Tau       &05:49:03.73 &$+$19:04:22.0 &A &  9.50     & 4.14      &2008/04/28& max     &SH,LH   & 50212\\
DY Per       &02:35:17.13 &$+$56:08:44.7 &C &  8.65     & 1.71      &2008/09/13& max     &SH,LH   & 50212\\
V517 Oph     &17:15:19.74 &$-$29:05:37.6 &C &  7.81     & 2.53      &2008/04/25& min[2.4]&SH,LH   & 50212\\
V CrA        &18:47:32.30 &$-$38:09:32.3 &AB&  5.66     & 2.46      &2005/09/14& max      &SL,LL   &     7\\
RZ Nor       &16:32:41.66 &$-$53:15:33.2 &A &  3.45     & 1.75      &2006/03/22& max      &SL,LL   &     7\\
RT Nor       &16:24:18.68 &$-$59:20:38.6 &A &  0.93     & 0.40      &2005/04/21& max     &SL,LL   &     7\\
RS Tel       &18:18:51.22 &$-$46:32:53.4 &A &  1.54     & 0.71      &2005/09/10& max     &SL,LL   &     7\\
V482 Cyg     &19:59:42.57 &$+$33:59:27.9 &A &  0.98     & 0.41      &2004/11/14& max     &SL,LL   &     7\\
MV Sgr       &18:44:31.97 &$-$20:57:12.8 &A &  0.60     & 1.57      &2005/04/18& max     &SL,SH,LH&  3362\\
RY Sgr       &19:16:32.76 &$-$33:31:20.4 &A & 77.20     &26.20      &2004/10/21& min[2.6]&SH,LH   &  3362\\
V854 Cen     &14:34:49.41 &$-$39:33:19.2 &AB& 23.00  & 7.82         &2007/09/07& min[5.5]&SL,SH,LH& 30077\\
UW Cen       &12:43:17.18 &$-$54:31:40.7 &A &  7.85  & 5.75         &2008/08/17& max     &SL,SH,LL& 40061\\
DY Cen       &13:25:34.08 &$-$54:14:43.1 &AB&  0.91  & 0.93         &2008/08/17& max       &SL,LL   & 40061\\
R CrB        &15:48:34.41 &$+$28:09:24.3 &A & 17.10  & 3.94         &2004/07/17& max       &SH      &    93\\
V348 Sgr     &18:40:19.93 &$-$22:54:29.3 &A &  5.53  & 3.00         &2006/10/22& min[2.7]&SL,LL   & 30380\\ 
HV 2671      &05:33 48.94 &$-$70:13:23.4 &LMC-RCB  &  $\dots$  & $\dots$         &2006/11/14&$\dots$ &SL,LL   & 30380\\
\enddata
\tablenotetext{a}{The first 18 RCB stars were observed with Spitzer by us
(Program \#50212) while the rest of stars were observed by other programs and
the data were retrieved from the Spitzer database (see text).} 
\tablenotetext{b}{Variability status during the Spitzer observations; max: the star was
observed at (or slightly below; e.g., $<$0.6$-$0.8 mag in V) maximum light. min: the star was observed during minimum light and the number
between brackets indicate the V magnitudes below maximum.} 
\end{deluxetable}

\clearpage

\begin{deluxetable}{lccccc}
\tabletypesize{\scriptsize}
\tablecaption{Optical and near infrared photometry$^{a}$ \label{tbl-2}}
\tablewidth{0pt}
\tablehead{
\colhead{RCB star} &  \colhead{Spitzer Obs. Date} & \colhead{IAC-80 Obs. Date} & \colhead{V/R/I$^{b}$} & \colhead{SAAO Obs. Date} &
\colhead{J/H/K/L}\\
\colhead{} &\colhead{yyyy/mm/dd} &\colhead{yyyy/mm/dd} &\colhead{} & \colhead{yyyy/mm/dd} &\colhead{}
}
\startdata
UV Cas        &2008/08/12 & 2008/08/11 & 10.73/9.86/9.01 & $\dots$& $\dots$\\
              &           & 2008/08/13 & 10.72/9.86/8.99 & & \\
S Aps         &2008/04/25 & $\dots$ & $\dots$ & 2008/04/24 & 7.78/7.16/6.67/5.31\\
SV Sge        &2008/05/26 & 2008/05/30 & 10.68/9.60/8.50 & $\dots$& $\dots$\\
Z UMi         &2008/10/20 & $\dots$ & $\dots$ & & \\
V1783 Sgr     &2008/04/25 & 2008/04/24 & 10.60/9.71/8.86 & 2008/04/29& 8.74/8.34/7.96/$\dots$ \\
              &           & 2008/05/04 & 10.71/9.82/8.93 & & \\
WX CrA        &2008/10/10 & $\dots$ & $\dots$ & & \\
V3795 Sgr     &2008/04/25 & 2008/04/24 & 11.57/10.94/10.27 &2008/04/29& 9.14/8.63/8.24/$\dots$  \\
              &           & 2008/05/04 & 11.57/10.94/10.29 & & \\
V1157 Sgr     &2008/06/02 & 2008/06/03 & 13.20/12.23/11.23 & $\dots$& $\dots$\\
Y Mus         &2008/04/25 & $\dots$ & $\dots$ & & \\
V739 Sgr      &2008/04/25 & 2008/04/24 & 12.41/11.33/10.33 & 2008/04/29& 10.11/9.49/8.70/$\dots$\\
              &           & 2008/05/04 & 12.42/11.66/10.45 & & \\
VZ Sgr        &2008/04/25 & $\dots$ & $\dots$ & & \\
U Aqr         &2008/06/29 & 2008/06/26 &11.40/10.74/10.23  & $\dots$& $\dots$\\
              &           & 2008/07/04 &11.50/10.81/10.30  & & \\
MACHOJ181933  &2008/04/25 & 2008/04/24 &13.93/12.78/11.64  & & \\
              &           & 2008/05/04 &13.97/12.86/11.73  & & \\
ES Aql        &2008/05/27 & 2008/05/30 &12.28/11.17/10.09  & & \\
FH Sct        &2008/06/02 & 2008/06/03 &13.00/12.90/11.99  & & \\
SU Tau        &2008/04/28 & 2008/04/23 & 9.79/ 9.15/ 8.56  &2008/03/19 & 7.72/7.28/6.74/4.96\\
DY Per        &2008/09/13 & $\dots$    & $\dots$              & & \\
V517 Oph      &2008/04/25 & 2008/04/24 &14.05/12.51/10.96  &2008/04/29 & 9.74/8.45/7.08/5.22\\
              &           & 2008/05/04 &13.73/12.25/10.74  & & \\
\enddata
\tablenotetext{a}{Optical and near infrared photometry for the RCB stars observed with Spitzer by us (Program \#50212).} 
\tablenotetext{b}{The VRI magnitude errors are estimated to be of the order of $\pm$0.15 mag (see text).} 
\end{deluxetable}

 \clearpage

\begin{deluxetable}{lccccccccccc}
\tabletypesize{\scriptsize}
\tablecaption{Blackbody fits to the RCB's SEDs \label{tbl-1}}
\tablewidth{0pt}
\tablehead{
\colhead{RCB star} & \colhead{T$_{star}$} & \colhead{T$_{BB1}$} & 
 \colhead{R$_{BB1}$} & \colhead{T$_{BB2}$} & \colhead{R$_{BB2}$}
& \colhead{T$_{BB1}$} & 
 \colhead{R$_{BB1}$} & \colhead{T$_{BB2}$} & \colhead{R$_{BB2}$} & \colhead{E(B-V)$^{a}$} & \colhead{$\Delta$T$^{b}$}\\
 \hline
 \colhead{} & \colhead{(K)} & \colhead{(K)} & \colhead{} & \colhead{(K)} & \colhead{} 
 & \colhead{(K)} &\colhead{} &\colhead{(K)} & \colhead{} & \colhead{} & \colhead{(days)}\\
\hline
\colhead{} & \colhead{} & \colhead{Spitzer} & \colhead{} & \colhead{} & \colhead{} 
 & \colhead{IRAS} &\colhead{} &\colhead{} & \colhead{} & \colhead{} & \colhead{}
}
\startdata
UV Cas       & 7200  & 510  &  0.03   & 180     & 0.001      & 800 & 0.28  & $\dots$ & $\dots$ & 0.90 & 25500\\
S Aps        & 4200  & 750  &  0.37   & $\dots$ & $\dots$ & 750 & 0.42  & $\dots$ & $\dots$    & 0.05 &  1400\\   
SV Sge       & 4200  & 565  &  0.05   & 350     & 0.024   & 720 & 0.15  & $\dots$ & $\dots$    & 0.72 &  2500\\ 
Z UMi        & 5200  & 710  &  0.43   & $\dots$ & $\dots$ & 850 & 0.95  & $\dots$ & $\dots$    & 0.00 &$\dots$\\
V1783 Sgr    & 5600  & 560  &  0.28   & $\dots$ & $\dots$ & 600 & 0.30  & $\dots$ & $\dots$    & 0.42 &$\dots$\\
WX CrA       & 4200  & 575  &  0.15   & 120     & 0.006   & 700 & 0.49  & $\dots$ & $\dots$    & 0.06 &  2000\\
V3795 Sgr    & 8000  & 610  &  0.31   & $\dots$ & $\dots$ & 720 & 0.54  & $\dots$ & $\dots$    & 0.79 &  6000\\
V1157 Sgr    & 4200  & 770  &  0.59   &  120    & 0.007   & 850 & 1.01  & $\dots$ & $\dots$    & 0.30 & $\dots$\\
Y Mus        & 7200  & 395  &  0.01   & $\dots$ & $\dots$ & 590 & 0.07  & $\dots$ & $\dots$    & 0.50 & 15300\\
V739 Sgr     & 5400  & 640  &  0.59   & 100     & 0.005   & 900 & 0.64  & 700     & 0.228      & 0.50 &$\dots$\\
VZ Sgr       & 7000  & 700  &  0.17   & 140     & 0.008   & 700 & 0.17  & 140     & 0.008      & 0.30 &  1300\\  
U Aqr        & 5000  & 475  &  0.23   & 140     & 0.021   & 560 & 0.37  & $\dots$ & $\dots$    & 0.05 &  1850\\
MACHOJ181933 & 4200  & 695  &  0.48   & 140     & 0.022   & $\dots$ & $\dots$ & $\dots$ & $\dots$& 0.50 &$\dots$ \\
ES Aql       & 4500  & 700  &  0.49   & $\dots$ & $\dots$ & 700 & 0.49  & $\dots$ & $\dots$    & 0.32 &$\dots$\\  
FH Sct       & 6250  & 540  &  0.10   & 140     & 0.002   & 390 & 0.04  & $\dots$ & $\dots$    & 1.00 &$\dots$\\ 
SU Tau       & 6500  & 635  &  0.45   & $\dots$ & $\dots$ & 635 & 0.50  & $\dots$ & $\dots$    & 0.50 &  1200\\ 
DY Per       & 3000  & 1400 &  0.31   & $\dots$ & $\dots$ &1400 & 0.31  & $\dots$ & $\dots$    & 0.48 &$\dots$\\
V517 Oph     & 4100  & 850  &  0.84   & $\dots$ & $\dots$ & 850 & 0.98  & $\dots$ & $\dots$    & 0.50 &$\dots$\\
V CrA$^{c}$        & 6500  & 550  &  0.38   & 150     & 0.020   &1600 & 0.23  & 900     & 0.370& 0.14 &   900\\  
RZ Nor       & 5000  & 700  &  0.53   & 320     & 0.035   & 700 & 0.53  & 300     & 0.040      & 0.50 &  1100\\
RT Nor       & 6700  & 320  &  0.01   & 130     & 0.001   & 500 & 0.11  & $\dots$ & $\dots$    & 0.39 &  1950\\
RS Tel       & 6750  & 720  &  0.25   & 130     & 0.005   & 620 & 0.22  & $\dots$ & $\dots$    & 0.17 &  1200\\
V482 Cyg     & 4800  & 500  &  0.03   & 100     & 0.001   & 650 & 0.09  & $\dots$ & $\dots$    & 0.50 &  3400\\ 
MV Sgr       &15400  & 1500 &  0.33   & 205     & 0.180   &1500 & 0.33  & 235     & 0.236      & 0.43 &  6900\\
RY Sgr       & 7200  & 675  &  0.20   &$\dots$  &$\dots$  & 870 & 0.76  & $\dots$ & $\dots$    & 0.00 &  1400\\
V854 Cen     & 6750  & 900  &  0.32   & 140     & 0.030   &1100 & 1.00  & $\dots$ & $\dots$    & 0.07 &   370\\
UW Cen$^{d}$ & 7500  & 630  &  0.44   & 120     & 0.013   & 630 & 0.44  & 150     & 0.033      & 0.32 &  1100\\
DY Cen       &19500  & 272  &  0.09   & $\dots$ & $\dots$ & 330 & 0.10  & $\dots$ & $\dots$    & 0.47 &  6400\\
R CrB        & 6750  & 950  &  0.30   & $\dots$ & $\dots$ & 680 & 0.20  & $\dots$ & $\dots$    & 0.00 &  1100\\     
V348 Sgr     &20000  & 707  &  0.63   & 100     & 0.035   & 707 & 0.63  & 100 & 0.035          & 0.45 &   560\\
HV 2671$^{e}$&20000  & 590  &  0.36   & 150     & 0.268   & $\dots$ & $\dots$ & $\dots$ & $\dots$ & 0.15 &$\dots$\\ 
\enddata
\tablenotetext{a}{See text for more details about the adopted E(B-V) values.}
\tablenotetext{b}{Inter-fade periods from Jurcsick et al. (1996) (see also the text for more details).}
\tablenotetext{c}{An additional 550 K blackbody with a covering factor of 0.37 is needed to fit the IRAS photometry.} 
\tablenotetext{d}{An additional very cool blackbody of 50 K with a covering factor of 0.05 is needed to fit the Spitzer data.} 
\tablenotetext{e}{An additional very cool blackbody of 40 K with a negligible covering factor is needed to fit the Spitzer data.}
\end{deluxetable}

\clearpage

\begin{figure}
\includegraphics[angle=0,scale=.60]{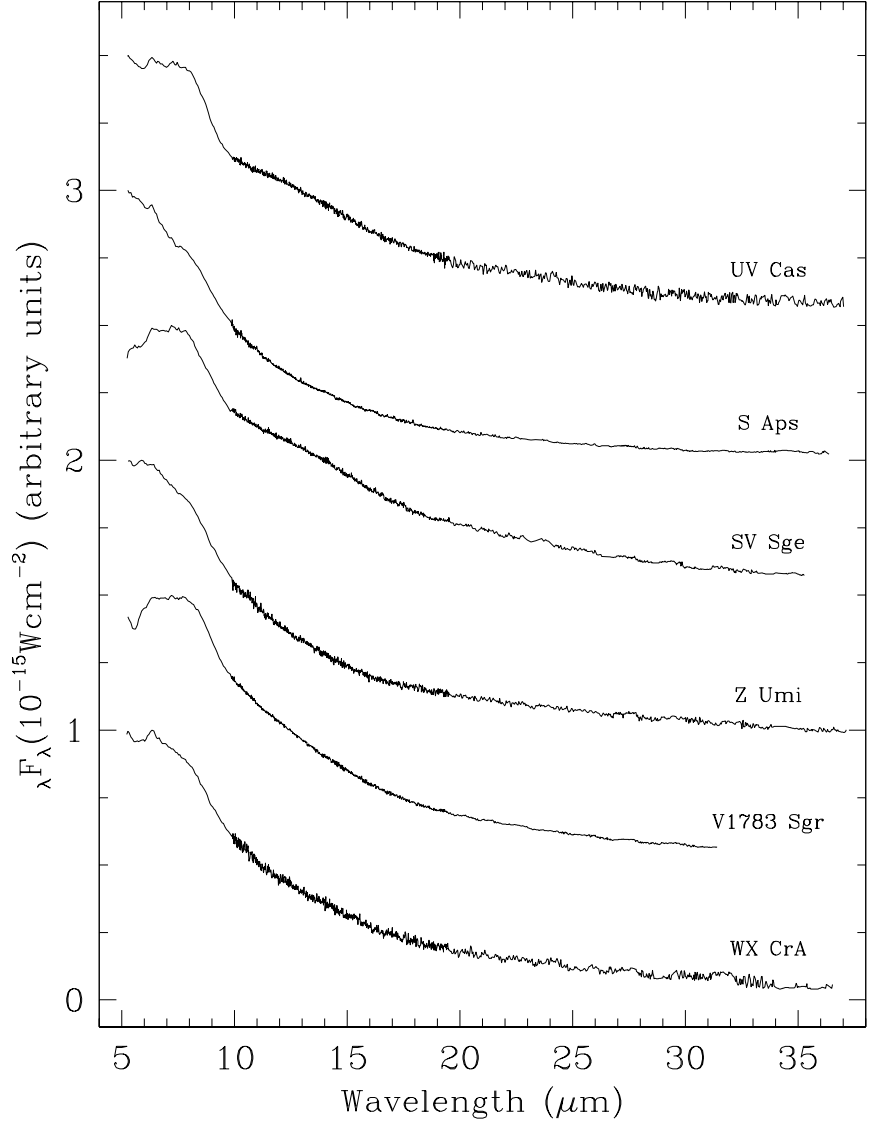}
\caption{Spitzer/IRS reduced spectra over the full wavelength range $\sim$5$-$37
$\mu$m for  (from top to bottom: UV Cas, S Aps, SV Sge, Z UMi,   V1783 Sgr and
WX CrA. Note that the spectra are normalized and displaced for clarity.
\label{fig1}}
\end{figure}

\clearpage

\begin{figure}
\includegraphics[angle=0,scale=.60]{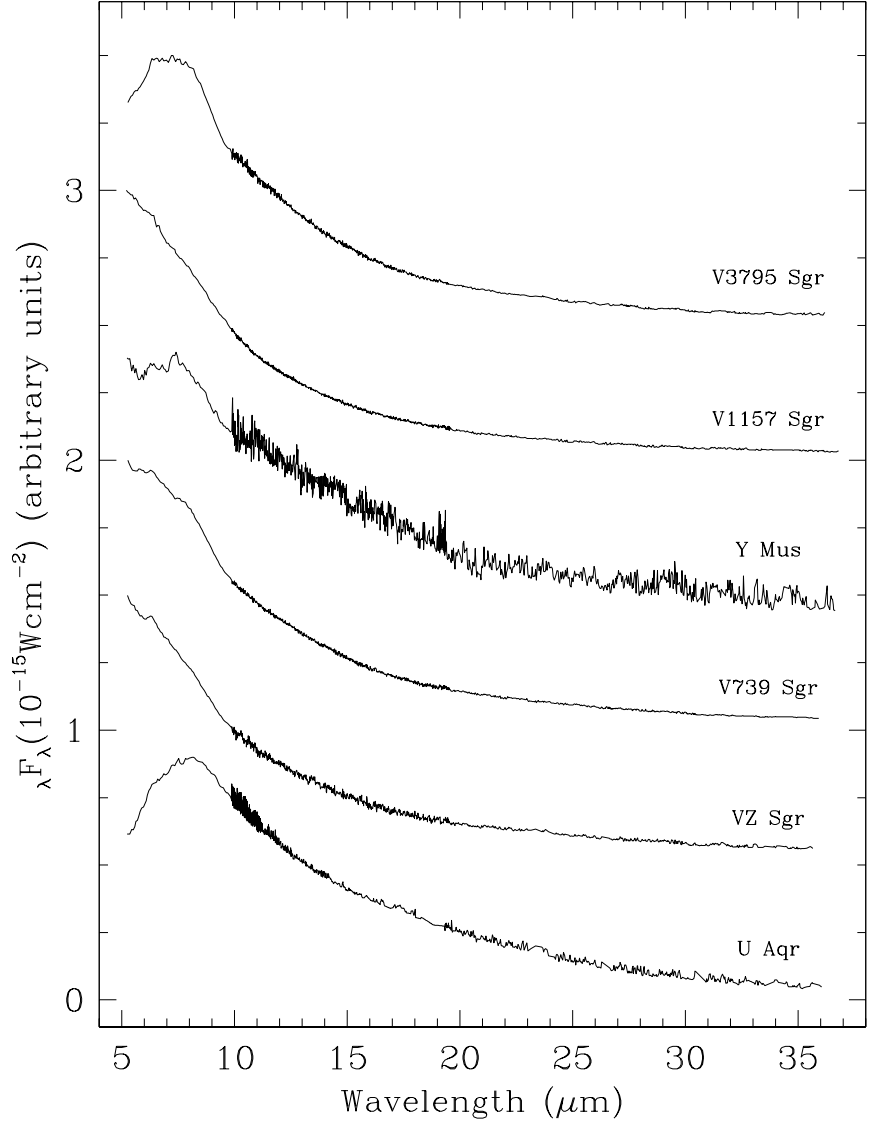}
\caption{Spitzer/IRS reduced spectra over the full wavelength range $\sim$5$-$37
$\mu$m for   V3795 Sgr, V1157 Sgr,  Y Mus, V739 Sgr, VZ Sgr, and U Aqr. Note
that the spectra are normalized and displaced for clarity.
\label{fig2}}
\end{figure}

\clearpage

\begin{figure}
\includegraphics[angle=0,scale=.60]{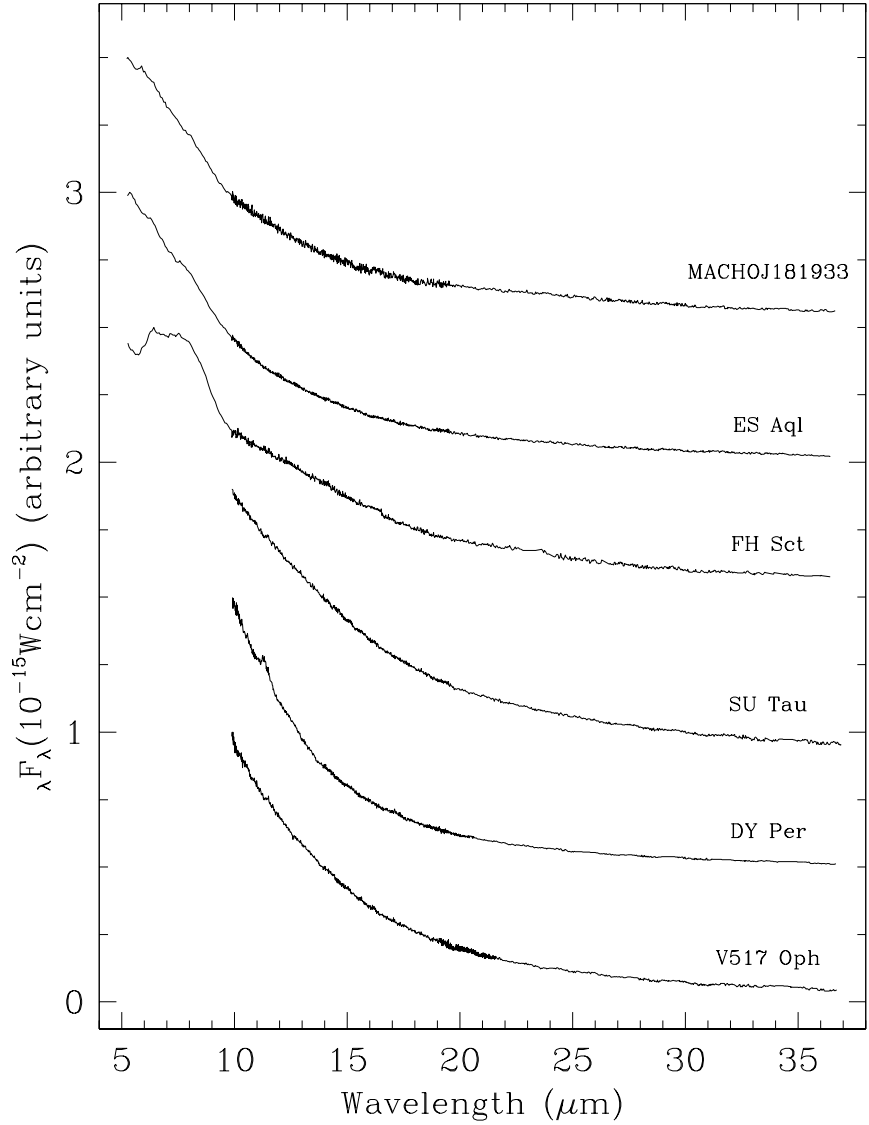}
\caption{Spitzer/IRS reduced spectra over the full wavelength range $\sim$5$-$37
$\mu$m for MACHOJ181933, ES Aql, FH Sct, SU Tau, DY Per, and V517 Oph. Note
that the spectra are normalized and displaced for clarity.
\label{fig3}}
\end{figure}

\clearpage

\begin{figure}
\includegraphics[angle=0,scale=.60]{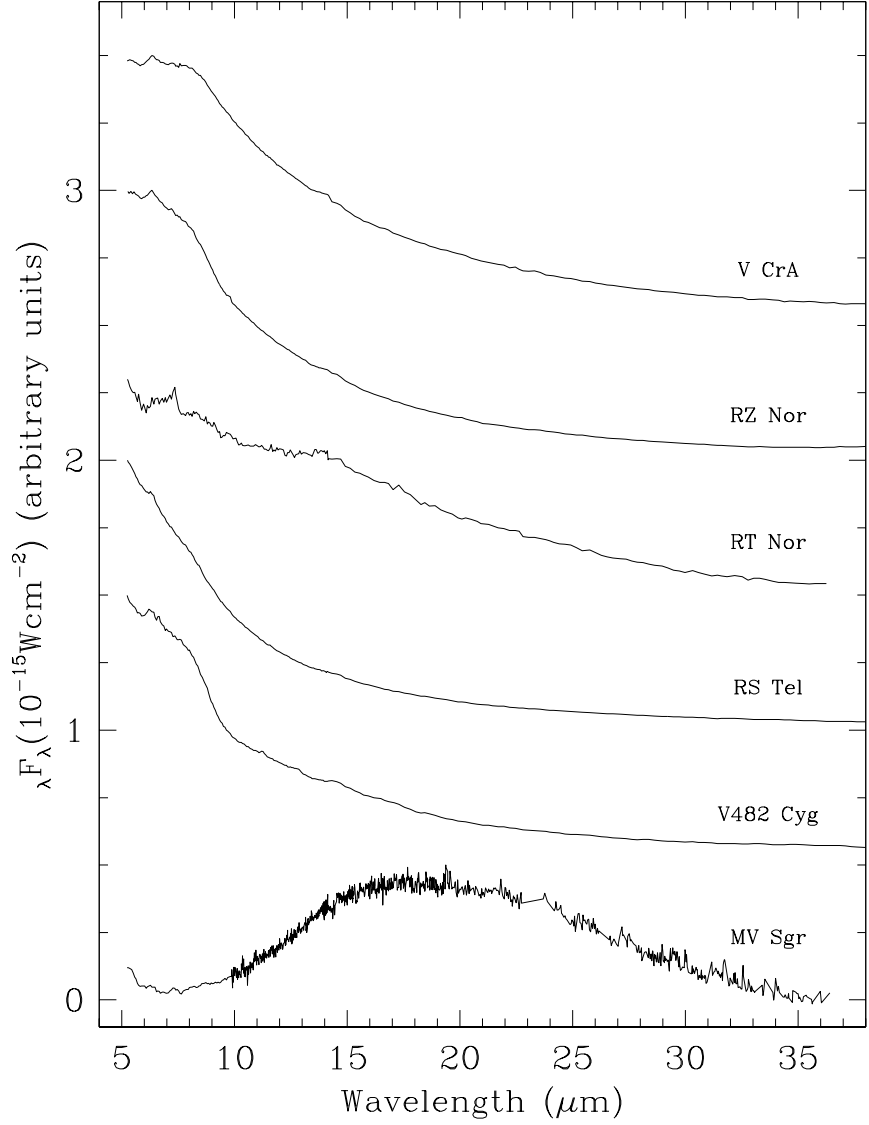}
\caption{Spitzer/IRS reduced spectra over the full wavelength range $\sim$5$-$37
$\mu$m for V CrA, RZ Nor, RT Nor, RS Tel, V482 Cyg, and MV Sgr. Note that the
spectra are normalized and displaced for clarity.
\label{fig4}}
\end{figure}

\clearpage

\begin{figure}
\includegraphics[angle=0,scale=.60]{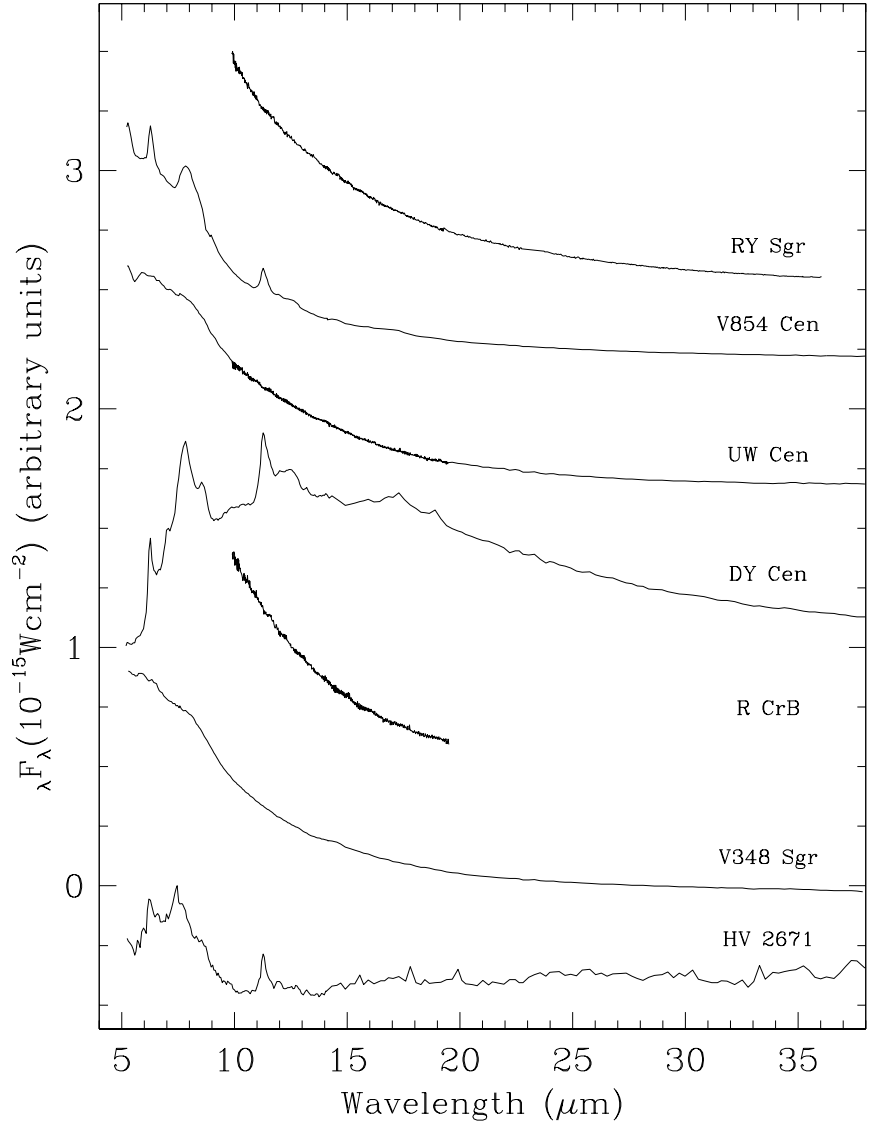}
\caption{Spitzer/IRS reduced spectra over the full wavelength range $\sim$5$-$37
$\mu$m for RY Sgr, V854 Cen, UW Cen, DY Cen, R CrB, V348 Sgr, and HV 2671. Note
that the spectra are normalized and displaced for clarity.
\label{fig5}}
\end{figure}

\clearpage

\begin{figure} 
\includegraphics[angle=0,scale=.60]{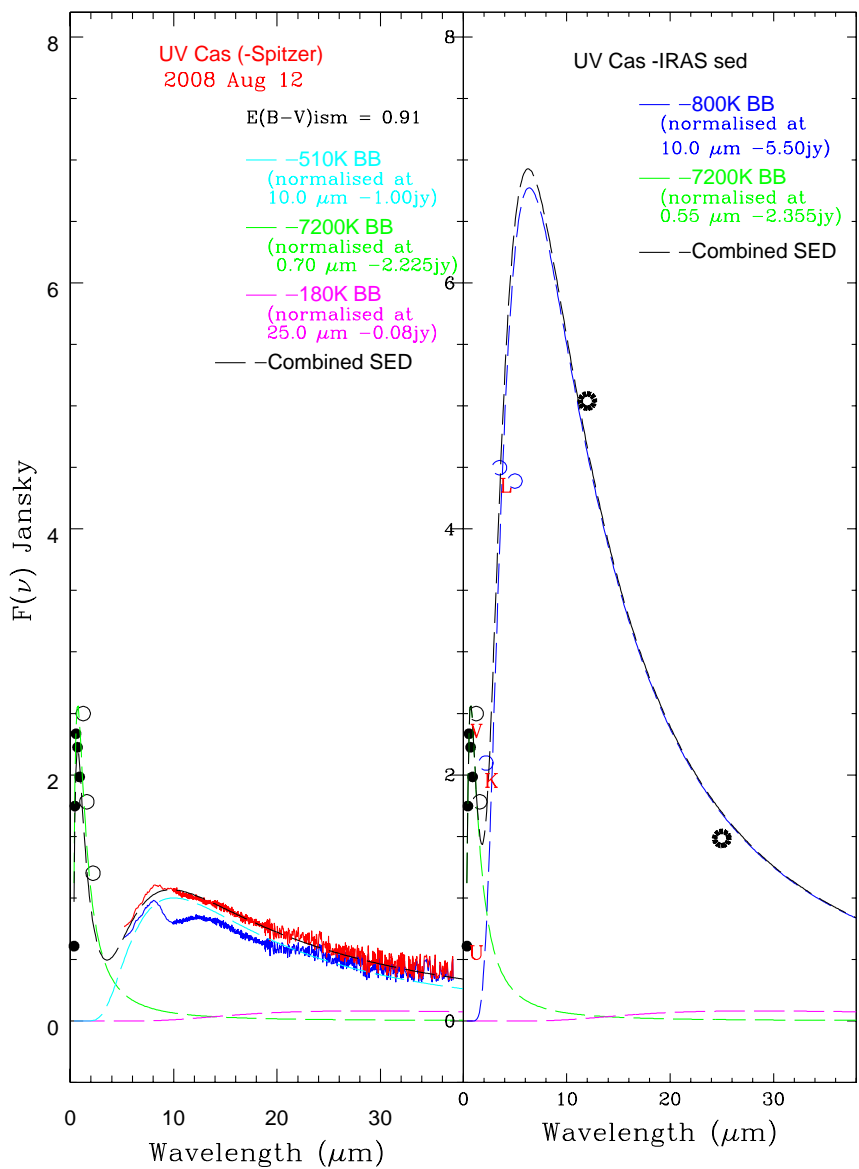}
\caption{Blackbody fits  for  UV  Cas. The lefthand panel shows a fit to the
stellar fluxes computed from reddening-corrected UBVRI (UBV: Fernie et al. 1972;
RI: this work) (black dots), 2MASS JHK (black open circles), and the Spitzer
spectrum (corrected for interstellar reddening, in red). The observed Spitzer
spectrum (in blue) and the blackbody temperatures are also shown. The righthand
panel shows a fit to the stellar UBVRI and 2MASS JH fluxes and IRAS 12 $\mu$m
and 25 $\mu$m fluxes with, in addition, the KLM fluxes (blue open circles)
estimated from Bogdanov et al. (2010) for the IRAS epoch. Selected UVKL
fluxes are labelled for convenience.
\label{fig6}}
\end{figure}

\clearpage

\begin{figure} 
\includegraphics[angle=0,scale=.60]{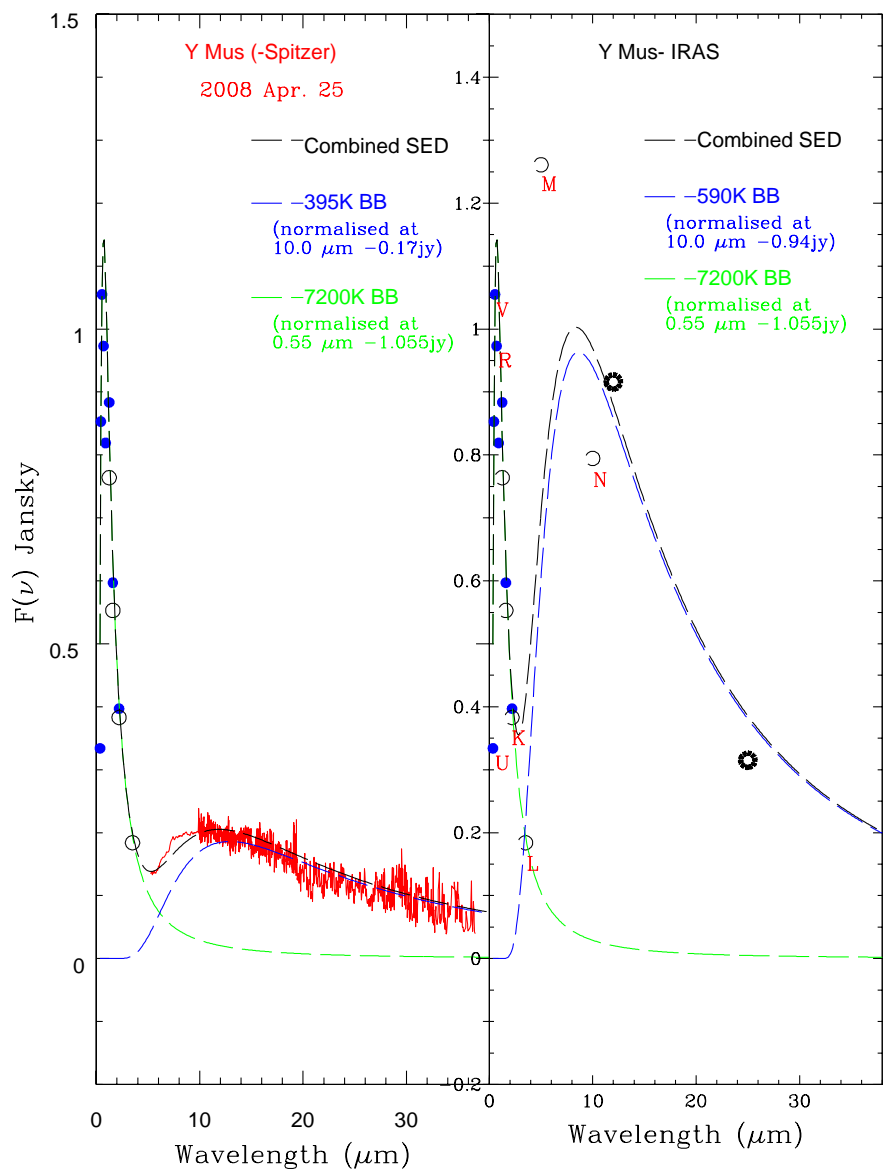}
\caption{Blackbody fits  for  Y Mus. The lefthand panel shows a fit to the
stellar fluxes computed from reddening-corrected  UBVRI (Kilkenny et al. 1985)
and JHK 2MASS photometry  (blue dots) and JHKL (Feast et al. 1997) (black open
circles), and the Spitzer spectrum (corrected for interstellar reddening, in
red). The observed Spitzer spectrum (in blue) and the blackbody temperatures are
also shown. The righthand panel shows a fit to the same stellar UBVRIJHK
photometric fluxes and IRAS 12 $\mu$m and 25 $\mu$m fluxes with, in addition,
fluxes at M and N from Kilkenny \& Whittet (1984).  Selected UVRKLMN fluxes
are labelled for convenience.
\label{fig7}} 
\end{figure}

\clearpage

\begin{figure} 
\includegraphics[angle=0,scale=.60]{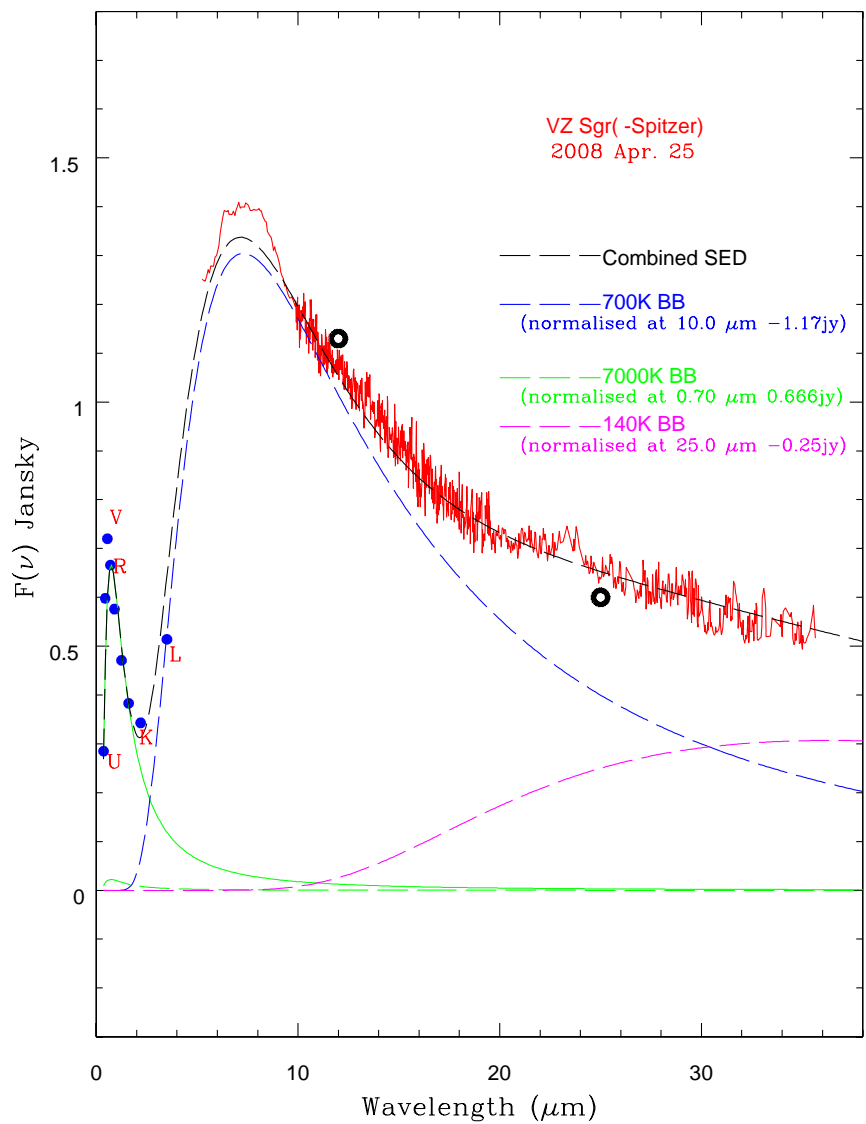} 
\caption{Blackbody fits for  VZ Sgr. Reddening-corrected ground-based photometry at UBVRIJHKL
(Kilkenny et al. 1985; Feast et al. 1997) and the Spitzer spectrum (in red) are
fitted with a stellar (7000 K) blackbody and dust blackbodies at 700 K and 140
K. Note that the correction for interstellar reddening is negligible for VZ Sgr
with E(B-V)=0.3 (see text). IRAS 12 $\mu$m and 25 $\mu$m fluxes are also shown
and straddle the Spitzer spectrum. Selected UVRKL fluxes are labelled for
convenience. \label{fig8}} 
\end{figure}

\clearpage

\begin{figure} 
\includegraphics[angle=0,scale=.60]{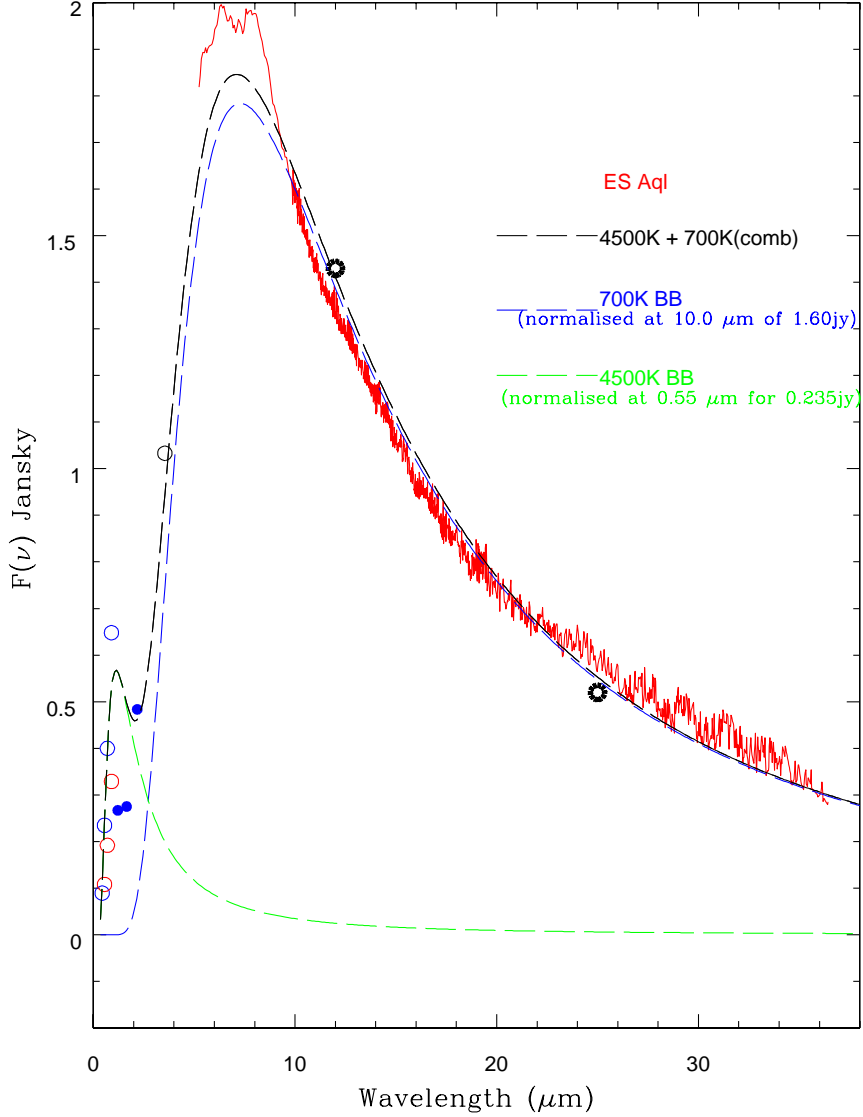} 
\caption{Blackbody fits for ES Aql.  Reddening-corrected VRI ground-based
photometry (red open circles) and the Spitzer spectrum (corrected for
interstellar reddening, in red) are fitted with a stellar (4500 K) blackbody and
a dust blackbody at 700 K.  IRAS 12 $\mu$m and 25 $\mu$m fluxes are shown and
straddle the Spitzer spectrum. Reddening-corrected BVRIJHKL fluxes (see text)
correspond to our VRI ground-based observations (red open circles, Table 2),
BVRI at maximum light (blue open circles), 2MASS JHK (blue dots) and L magnitude
(black open circle).\label{fig9}} 
\end{figure}

\clearpage

\begin{figure} 
\includegraphics[angle=0,scale=.60]{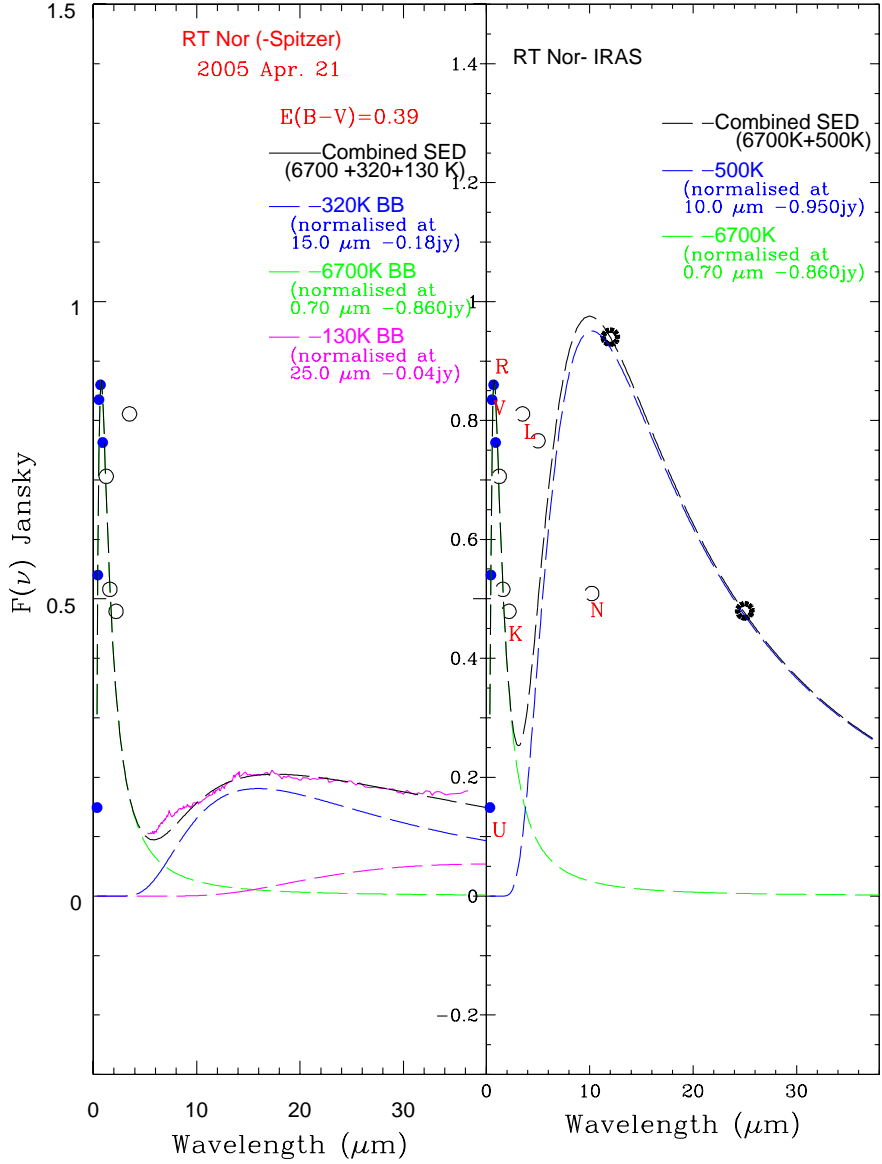} 
\caption{Blackbody fits  for  RT Nor. The lefthand panel shows a fit to the
stellar fluxes computed from reddening-corrected  UBVRI (Kilkenny et al. 1985)
(blue dots) and JHKL (Feast et al. 1997) (open circles) and the Spitzer spectrum
(in red) with the blackbody temperatures shown on the panel. Note that the
correction for interstellar reddening is negligible for RT Nor with E(B-V)=0.39
(see text). The righthand panel shows a fit to the same stellar UBVRIJHKL
photometric  fluxes and IRAS 12 $\mu$m and 25 $\mu$m fluxes with, in addition,
fluxes at M and N from Kilkenny \& Whittet (1984).  Selected UVRKLN fluxes
are labelled for convenience.
\label{fig10}} 
\end{figure}

\clearpage

\begin{figure} 
\includegraphics[angle=0,scale=.60]{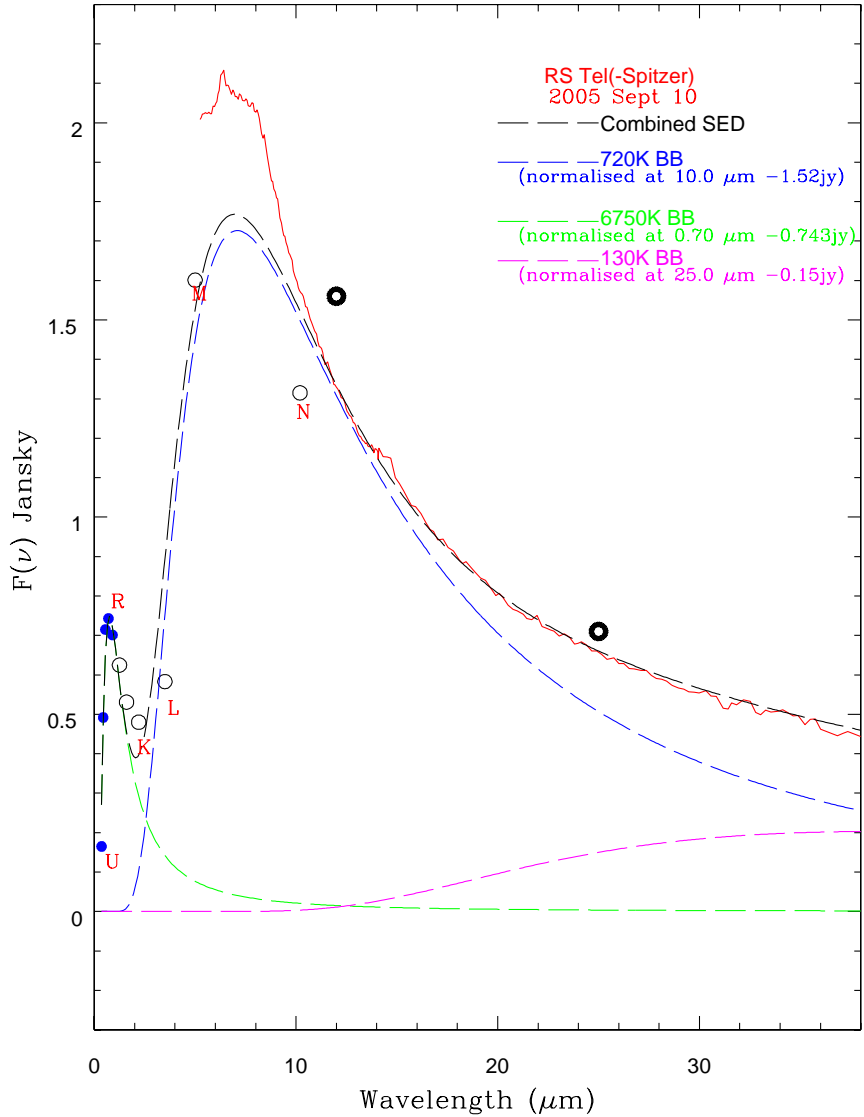} 
\caption{Blackbody fits for RS Tel. Ground-based reddening-corrected photometry
at UBVRIJHKL (Kilkenny et al. 1985; Glass 1978; see text) and the Spitzer
spectrum (in red) are fitted with a stellar (6750 K) blackbody and dust
blackbodies at 720 K and 130 K.  Note that the correction for interstellar
reddening is negligible for RS Tel with E(B-V)=0.17 (see text). IRAS 12 $\mu$m
and 25 $\mu$m fluxes and MN from Kilkenny \& Whittet (1984) are also shown.
Selected URKLMN fluxes are labelled for convenience.
\label{fig11}} 
\end{figure}

\clearpage

\begin{figure} 
\begin{center}$
\begin{array}{cc}
\includegraphics[angle=0,scale=.40]{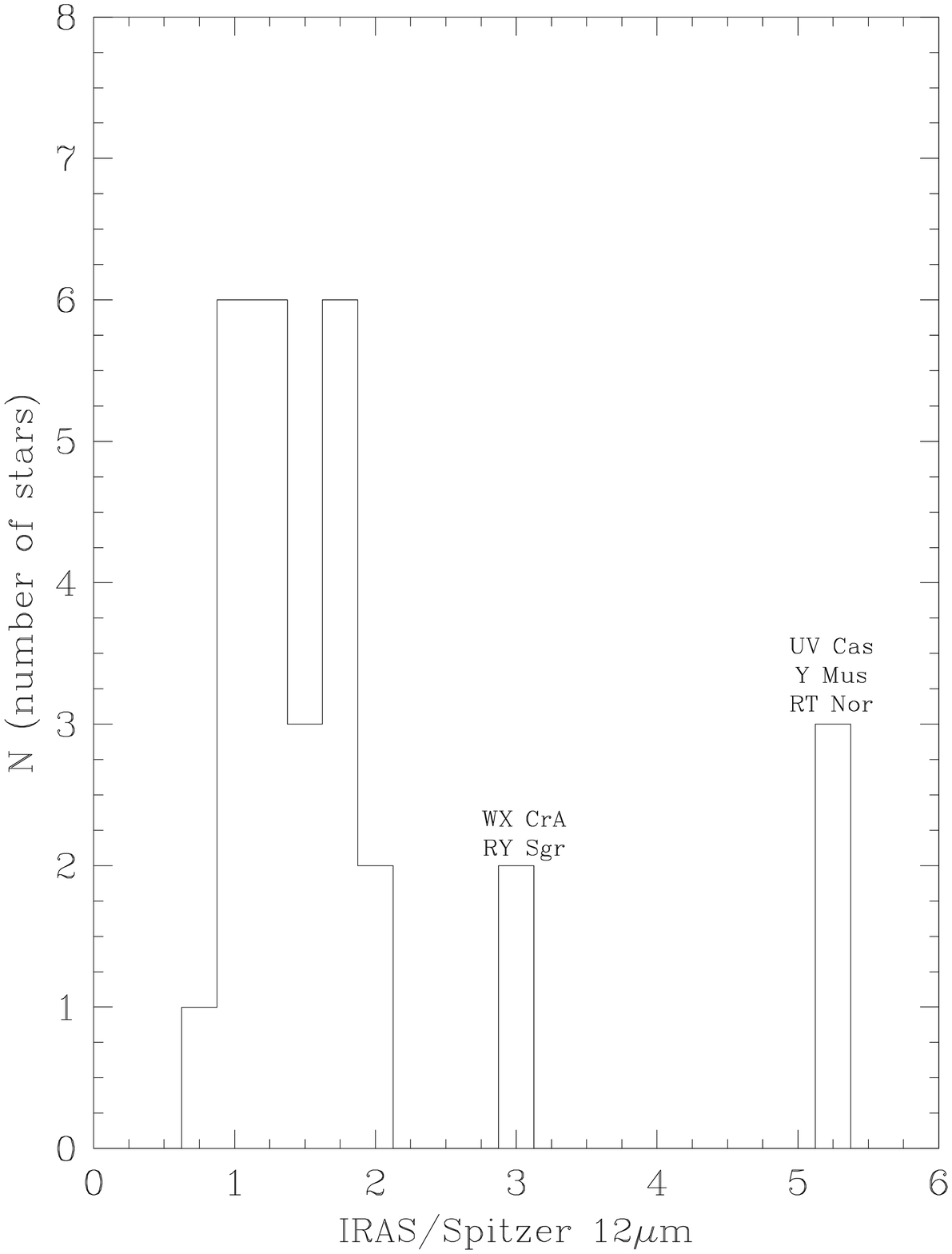} &
\includegraphics[angle=0,scale=.40]{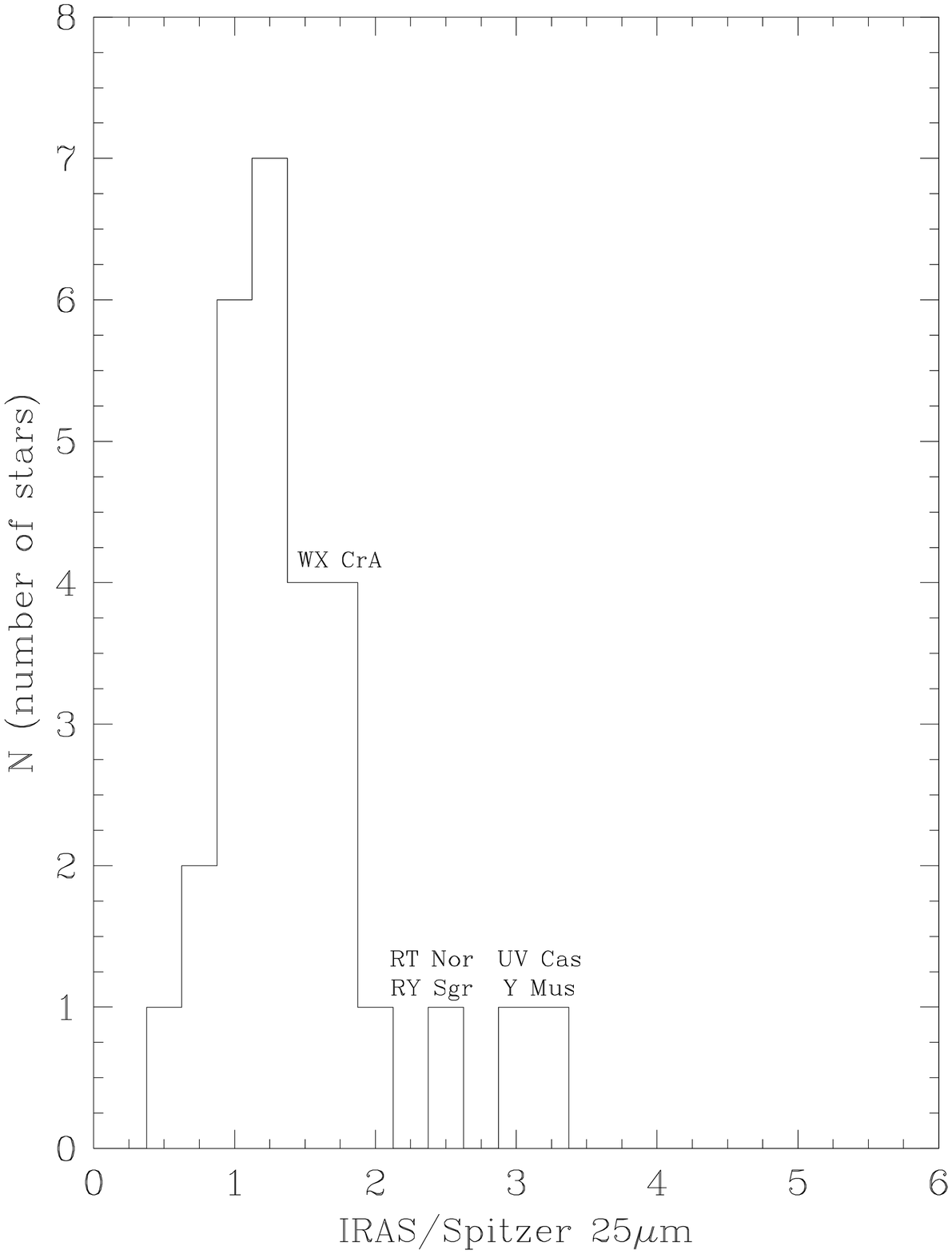}
\end{array}$
\end{center}
\caption{Histograms of the IRAS/Spitzer flux ratios. Left panel - fluxes at
12 $\mu$m. Right panel - fluxes at 25 $\mu$m. Fluxes are uncorrected for
interstellar reddening. Color corrections have not been applied to the IRAS
fluxes. Five outliers are identified.
\label{fig12}} 
\end{figure}

\clearpage

\begin{figure} 
\includegraphics[angle=0,scale=.60]{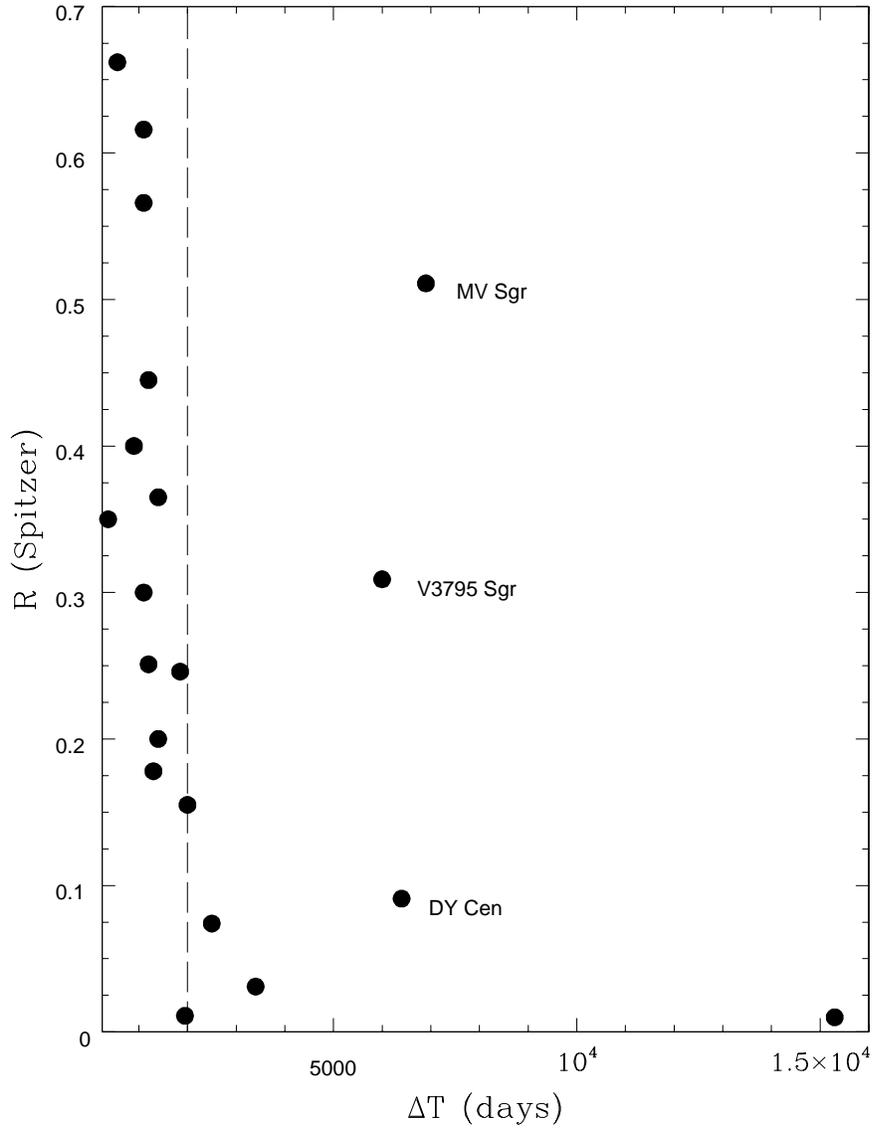} 
\caption{Spitzer Covering factor R versus the inter-fade period $\Delta$T. The
dashed vertical line marks the point $\Delta$T=2000 days discussed in the text.
Note that UV Cas with the longest $\Delta$T=25500 days and a very low R=0.035 is
not shown for clarity. \label{fig13}} 
\end{figure}

\clearpage

\begin{figure} 
\includegraphics[angle=0,scale=.60]{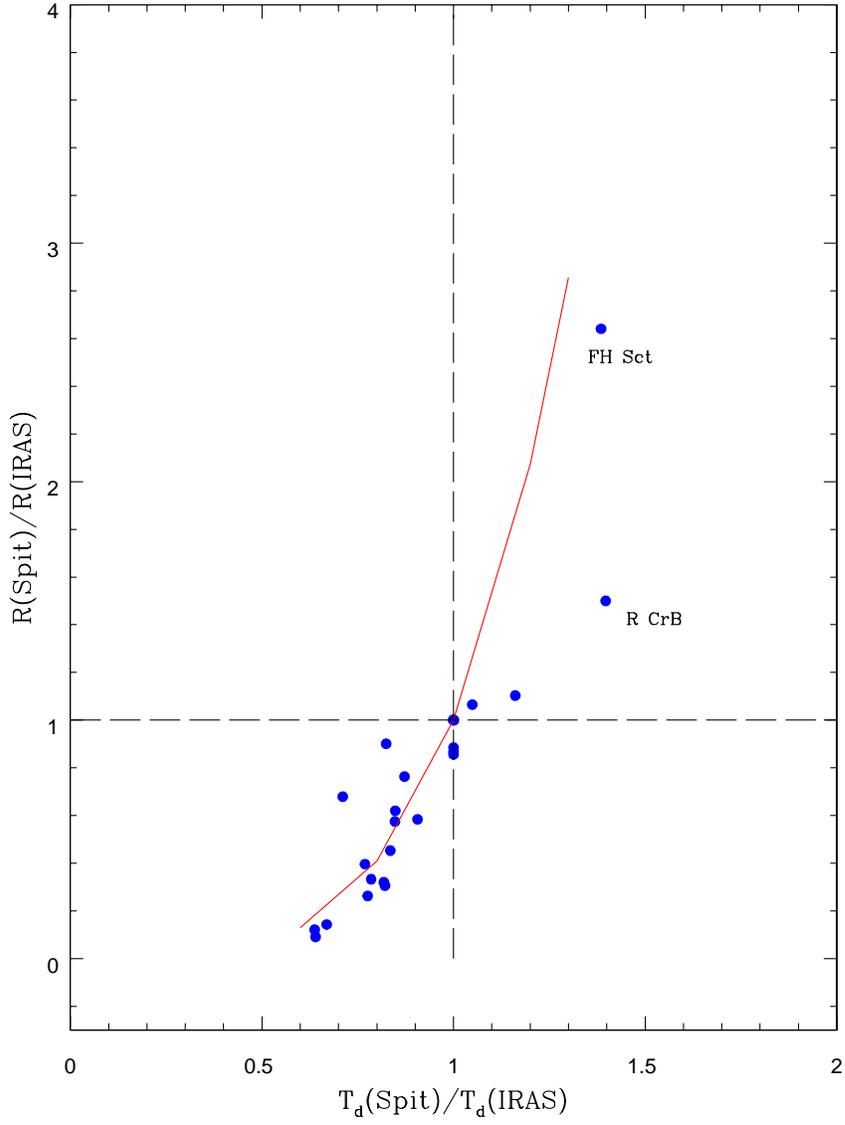} 
\caption{The ratio of the Spitzer to IRAS blackbody dust temperatures  versus
the ratio of the Spitzer to IRAS covering factors. The solid line
is the simple prediction discussed in the text.
\label{fig14}} 
\end{figure}

\clearpage

\begin{figure} 
\includegraphics[angle=0,scale=.60]{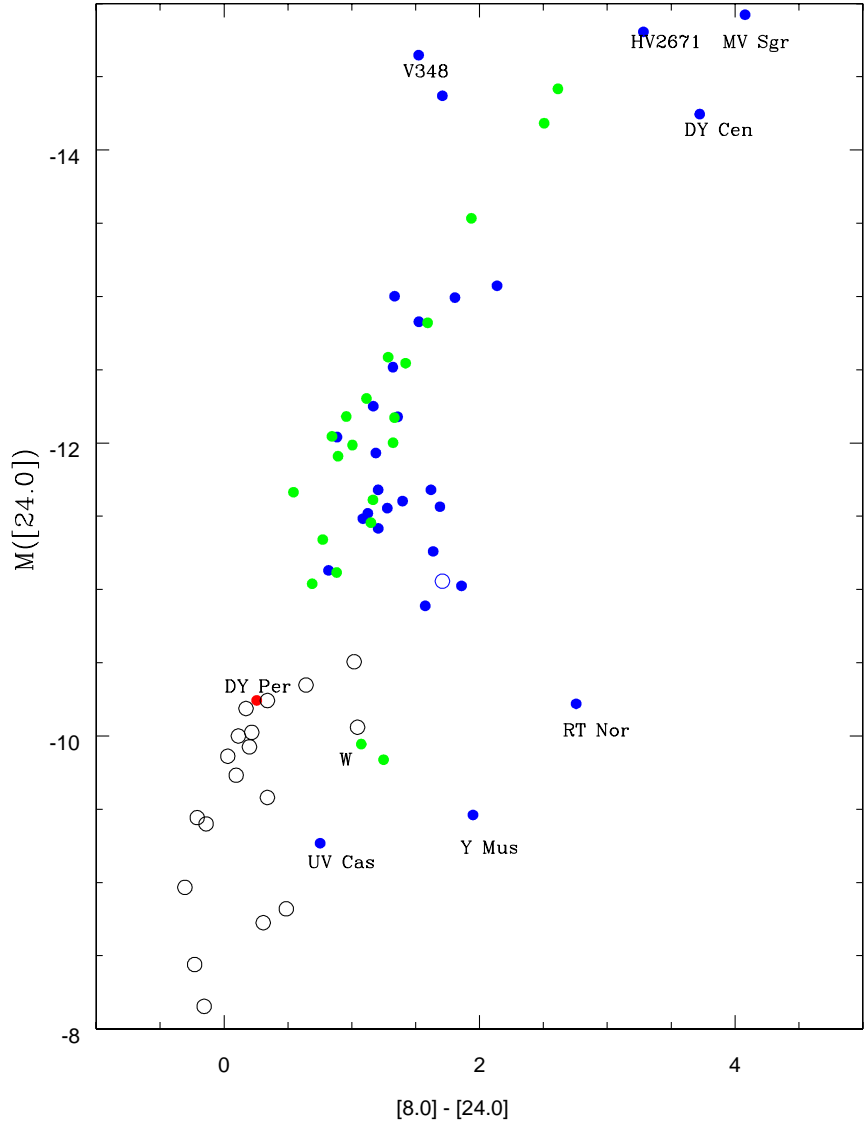} 
\caption{[24.0]$\mu$m absolute magnitude versus the [8.0]$-$[24.0] color index
for our sample of Galactic RCBs (blue dots) in comparison with LMC RCBs (green
dots) and LMC DY Per-like stars (black open circles). Hot RCBs (V348 Sgr, DY
Cen, MV Sgr and HV 2671) are located towards the [24.0]$\mu$m high luminosity
end and are labelled. The three RCB stars (UV Cas, Y Mus, and RT Nor) showing
remarkably low [24.0]$\mu$m magnitudes are also labelled. Note that DY Per
itself and W Men are also marked with a red dot and the letter `W',
respectively. \label{fig15}} 
\end{figure}


\begin{thebibliography}{}
\bibitem[Alcock et al.(2001)]{alc01} Alcock, C., Allsman, R. A., Alves, D. R. et al. 2001, ApJ, 554, 298
\bibitem[Alksnis et al.(2009)]{alk09} Alksnis, A., Larionov, V.M., Smirnova, O., Arkharov, A.A., Konstantinova, T.S., Larionova, L.V., \& Shenavrin, V.I. 2009, BaltA, 18, 53
\bibitem[Asplund et al.(1997)]{aspl97} Asplund, M.,Gustafsson, B., Kiselman, D.,  \& Eriksson, K. 1997, A\&A, 318, 521
\bibitem[Asplund et al.(1998)]{aspl98} Asplund, M., Gustafsson, B., Rao, N.K., \& Lambert, D.L. 1998, A\&A, 332, 651
\bibitem[Asplund et al.(2000)]{aspl00} Asplund, M.,Gustafsson, B., Lambert, D. L., \& Rao, N. K. 2000, A\&A, 353, 287
\bibitem[Beichman(1988)]{bei88} Beichman, C.A., Neugebauer, G., Habing, H.J., Clegg, P.E., \& Chester, T.J. 1988, Infrared Astronomical Satellite (IRAS) Catalogs and Atlases, Vol. 1: Explanatory Supplement (Washington:NASA)
\bibitem[Benson et al.(1994)]{ben94} Benson, P.J., Clayton, G.C., Garnavich, P., \& Szkody, P. 1994, AJ, 108, 247 
\bibitem[Bergeat et al.(1999)]{ber99} Bergeat, J., Knapik, A., \& Rutily, B.D. 1999, A\&A, 342, 773
\bibitem[Bogdanov et al.(2010)]{bog10} Bogdanov, M.B., Taranov, O.G., \& Shenavrin, V.I. 2010, Astr. Reports, 54, 620
\bibitem[Bond et al.(1979)]{bon79} Bond, H.E., Luck, R.E., \& Newman, M.J. 1979, ApJ, 233, 205
\bibitem[Bright et al.(2011)]{bri11} Bright, S. N., Chesneau, O., Clayton, G. C., De Marco, O., Leão, I. C., Nordhaus, J., Gallagher, J. S. 2011, MNRAS (in press; astro-ph/1102.4147)
\bibitem[Cardelli et al.(1989)]{car89} Cardelli, J. A., Clayton, G. C., \& Mathis, J. S. 1989, ApJ, 345, 245
\bibitem[Carter(1990)]{car90} Carter, B.S. 1990, MNRAS, 242, 1
\bibitem[Chiar \& Tielens(2006)]{chiar06} Chiar, J. E., \& Tielens, A. G. G. M. 2006, ApJ, 637, 774
\bibitem[Clayton et al.(1992)]{clay92} Clayton, G. C., Whitney, B. A., Stanford, S. A., Drilling, J. S. 1992, ApJ, 397, 652
\bibitem[Clayton et al.(1995)]{clay95} Clayton, G.C., Kelly, D.M., Lacy, J.H., Little-Marenin, I.R., Feldman, P.R., \& Bernath, P.F. 1995, AJ, 109, 2096
\bibitem[Clayton(1996)]{clay96} Clayton, G. C. 1996, PASP, 108, 225
\bibitem[Clayton et al.(1997)]{clay97} Clayton,G. C., Bjorkman, K. S., Nordsieck,,K. H., Zellner, E. B., Schulte-Ladbeck, R. E. 1997, ApJ, 476, 870
\bibitem[Clayton et al.(1999)]{clay99} Clayton, G.C., Kerber, F., Gordon, K.D., Lawson, W.A., Wolff, M.J., Pollacco, D.L., \& Furlan, E. 1999, ApJ, 517, L143
\bibitem[Clayton et al.(2002)]{clay02} Clayton, G. C., Hammond, D., Lawless, J., Kilkenny, D., Evans, T. L., Mattei, J., Landolt, A. U. 2002, PASP, 114, 846
\bibitem[Clayton et al.(2005)]{clay05} Clayton, G.C., Herwig, F., Geballe. T.R., Asplund, M., Tenenbaum, E.D., Engelbracht, C.W., \& Gordon, K.D. 2005, ApJ, 623, L141
\bibitem[Clayton et al.(2007)]{clay07} Clayton, G.C., Geballe, T.R., Herwig, F., Fryer, C. \& Asplund, M. 2007, ApJ, 662, 1220
\bibitem[Clayton et al.(2011)]{clay11} Clayton, G.C., De Marco, O., Whitney, B. A., Babler, B., Gallagher, J. S., Nordhaus, J., Speck, A. K., Wolff, M. J., Freeman, W. R., Camp, K. A., Lawson, W. A., Roman-Duval, J., Misselt, K. A., Meade, M., Sonneborn, G., Matsuura, M., Meixner, M. 2011, AJ (in press)
\bibitem[Colangeli et al.(1995)]{col95} Colangeli, L., Mennella, V., Palumbo, P., Rotundi, A., Bussoletti, E. 1995, A\&AS, 113, 561
\bibitem[Cottrell \& Lambert(1982)]{cott82} Cottrell, P.L., \& Lambert, D.L. 1982, The Observatory, 102, 149
\bibitem[Cox et al.(2011)]{cox11} Cox, N. L. J., Garc\'{\i}a-Hern\'andez, D. A., Garc\'{\i}a-Lario, P., Manchado, A. 2011, AJ, 141, 111
\bibitem[Crause et al.(2007)]{crau07} Crause, L.A., Lawson, W.A., \& Henden, A.A. 2007, MNRAS, 375, 301
\bibitem[Cutri et al.(2003)]{cut03} Cutri, R. M., et al., 2003, The 2MASS All-Sky Point Source Catalog, Vizier Online Catalog, 2246, 0 
\bibitem[Danziger (1965)]{dan65} Danziger, I.J., 1965, MNRAS, 130, 199 
\bibitem[de Laverny et al. (2004)]{delavern04} de Laverny, P., \& M\'{e}karnia, D. 2004, A\&A, 428, L13
\bibitem[De Marco et al.(2002)]{demarc02} De Marco, O., Clayton, G. C., Herwig, F., Pollacco, D. L., Clark, J. S., Kilkenny, D. 2002, AJ, 123, 3387
\bibitem[Engelbracht et al.(2007)]{eng07} Engelbracht, C. W. et al. 2007, PASP, 119, 994
\bibitem[Feast(1979)]{feast79} Feast, M.W. 1979, in IAU Colloq. 46, Changing Trends in Variable Star Research, ed. F.M. Bateson, J. Smak, \& I.M. Urch (Hamilton: Univ.  Waikato) 246 
\bibitem[Feast(1986)]{feast86} Feast, M.W., in  IAU Colloq. 87,  Hydrogen Deficient Stars and Related Objects, ed. K. Hunger, D. Sch\"{o}nberner, \& N.K. Rao (Dordrecht: Reidel) 151
\bibitem[Feast(1996)]{feast96} Feast, M.W. 1996, in ASP Conf. Ser. 96, Hydrogen Deficient Stars, ed. C.S. Jeffery \& U. Heber (San Francisco:ASP) 3
\bibitem[Feast(1997)]{feast97} Feast, M.W. 1997, MNRAS, 285, 339
\bibitem[Feast \& Glass(1973)]{feast73} Feast, M.W., \& Glass, I.S. 1973, MNRAS, 161, 293 
\bibitem[Feast et al.(1997)]{feast97} Feast, M. W., Carter, B. S., Roberts, G., Marang, F., Catchpole, R. W. 1997, MNRAS, 285, 317
\bibitem[Fernie et al.(1972)]{fer72} Fernie, J.D., Sherwood, V., \& DuPuy, D.L. 1972, ApJ, 172, 383
\bibitem[Fitzgerald(1968)]{fit68} Fitzgerald, M.P. 1968, AJ, 73, 683
\bibitem[Forrest, Gillett \& Stein(1972)]{forr72} Forrest, W. J., Gillett, F. C., \& Stein, W. A. 1972, ApJ, 178, L17
\bibitem[Garc\'{\i}a-Hern\'andez et al.(2007)]{gh07} Garc\'{\i}a-Hern\'andez D. A., Perea-Calder\'on, J. V., Bobrowsky, M., Garc\'{\i}a-Lario, P. 2007, ApJ, 666, L33
\bibitem[Garc\'{\i}a-Hern\'andez et al.(2009)]{gh09} Garc\'{\i}a-Hern\'andez D. A., Perea-Calder\'on, J. V., Engels, D., Garc\'{\i}a-Lario, P. 2009 in ``Asymmetrical Planetary Nebulae IV", p. 325
\bibitem[Garc\'{\i}a-Hern\'andez et al.(2009)]{gh09} Garc\'{\i}a-Hern\'andez D. A., Hinkle, K. H., Lambert, D. L., Eriksson, K. 2009, ApJ, 696, 1733
\bibitem[Garc\'{\i}a-Hern\'andez et al.(2010)]{gh10} Garc\'{\i}a-Hern\'andez D. A., Lambert, D. L., Rao, N. K., Hinkle, K. H., Eriksson, K. 2010, ApJ, 714, 144
\bibitem[Garc\'{\i}a-Hern\'andez et al.(2011)]{gh11} Garc\'{\i}a-Hern\'andez D. A., Rao, N. K., \& Lambert, D. L. 2011, ApJ, 729, 126 
\bibitem[Gaustad et al.(1988)]{gau88} Gaustad, J.E., Stein, W.A., Forrest, W.J., \& Pipher, J.L. 1988, PASP, 100, 388
\bibitem[Geballe et al.(2009)]{geba09} Geballe, T. R., Rao, N. K., \& Clayton, G. C. 2009, ApJ, 698, 735
\bibitem[Gillett et al.(1986)]{gill86} Gillett, F. C., Backman, D. E., Beichman, C., \& Neugebauer, G. 1986, ApJ, 310, 842
\bibitem[Glass(1978)]{gla78} Glass, I.S. 1978, MNRAS, 185, 23
\bibitem[Goldsmith et al.(1990)]{gol90} Goldsmith, M.J., Evans, A., Albinson, J.S., \& Bode, M.F. 1990, MNRAS, 245, 119
\bibitem[Goswami et al. (1997)]{gos97} Goswami, A., Rao, N.K., \& Lambert, D.L. 1997, PASP, 109, 796
\bibitem[Heck et al.(1985)]{hec85} Heck, A., Houziaux, L., Manfroid, J., Jones, D.H.P., \& Andrews, P.J. 1985, A\&AS, 61, 375
\bibitem[Herbig(1969)]{herb69} Herbig, G. H. 1969, M\'em. Soc. Roy. Sc Liege, 19, 13
\bibitem[Higdon et al.(2004)]{hig04} Higdon, S. J. U. et al. 2004, PASP, 116, 975
\bibitem[Hoffleit(1959)]{hoff59} Hoffleit, D. 1959, AJ, 64, 241
\bibitem[Houck et al.(2004)]{hou04} Houck, J. R. et al. 2004, ApJS, 154, 18
\bibitem[Jeffery et al.(1988)]{jeff88} Jeffery, C.S., Heber, U., Hill, P.W., \& Pollacco, D. 1988, MNRAS, 231, 175
\bibitem[Jeffery \& Heber(1993)]{jeff93} Jeffery, C. S., \& Heber, U. 1993, A\&A, 270, 167
\bibitem[Jeffery(1995)]{jeff95} Jeffery, C.S. 1995, A\&A, 297, 779
\bibitem[Jeffery et al.(2011)]{jeff11} Jeffery, C. S., Karakas, A. I., \& Saio, H. 2011, MNRAS (in press)
\bibitem[Jurcsik(1996)]{jur96} Jurcsik, J. 1996, Acta Astron., 46, 325
\bibitem[Kilkenny \& Whittet(1984)]{kilk84} Kilkenny, D., \& Whittet, D.C.B. 1984, MNRAS, 208, 25
\bibitem[Kilkenny et al.(1985)]{kilk85} Kilkenny, D., Coulson, I.M., Laing, J.D., Spencer Jones, J. \& Engelbrecht, C. 1985, SAAO Circ., 9, 87
\bibitem[Kilkenny et al.(1992)]{kilk92} Kilkenny, D., Lloyd Evans, T., Bateson, F. M., Jones, A. F., Lawson, W. A. 1992, The Observatory, 112, 158
\bibitem[Kipper \& Klochkova(2006)]{kip06} Kipper, T., \& Klochkova, V.G. 2006, BaltA, 15, 531
\bibitem[Kwok(1976)]{kwo76} Kwok, S. 1976, JRASC, 70, 49
\bibitem[Kwok(2007)]{kwo07} Kwok, S. 2007, Physics and Chemistry of the Interstellar Medium, (Sausalito:University Science Books)
\bibitem[Lambert \& Rao(1994)]{lam94} Lambert, D. L., \& Rao, N. K. 1994, JApA, 15, 47
\bibitem[Lambert et al.(2001)]{lam01} Lambert, D. L., Rao, N. K., Pandey, G., Ivans, I. I. 2001, ApJ, 555, 925
\bibitem[Lawson et al.(1990)]{law90} Lawson, W.A., Cottrell, P.L., Kilmartin, P.M., \& Gilmore, A.C. 1990, MNRAS, 247, 91
\bibitem[Lawson \& Cottrell(1997)]{laws97} Lawson, W. A., \& Cottrell, P.L. 1997, MNRAS, 285, 266
\bibitem[Lawson et al.(1999)]{law99} Lawson, W.A. et al. 1999, AJ, 117, 3007
\bibitem[Le\~ao et al.(2007)]{leao07} Le\~ao, I. C., de Laverny, P., Chesneau, O., M\'ekarnia, D., de Medeiros, J. R. 2007, A\&A, 466, L1
\bibitem[Lloyd Evans et al.(1991)]{lloy91} Lloyd Evans, T., Kilkenny, D., van Wyk, F. 1991, The Observatory, 111, 244
\bibitem[Marang et al.(1990)]{mar90} Marang, F., Kilkenny, D., Menzies, J.W., \& Spencer Jones, J.H. 1990, SAAO Circ., 14, 1
\bibitem[Pandey \& Lambert(2011)]{pan11} Pandey, G., \& Lambert, D. L. 2011, ApJ, 727, 122
\bibitem[Pollacco \& Hill(1991)]{poll91a} Pollacco, D.L., \& Hill, P.W. 1991, MNRAS, 248, 572
\bibitem[Pollacco et al.(1990)]{poll90} Pollacco, D. L., Hill, P. W., \& Tadhunter, C. N., 1990, MNRAS, 245, 204
\bibitem[Pollacco et al.(1991)]{poll91b} Pollacco, D.L., Hill, P.W., Houziaux, L., \& Manfroid, J. 1991, MNRAS, 248, 1p
\bibitem[Pugach (1977)]{pug77} Pugach, A.E. 1977, IBVS, 1277, 1
\bibitem[Rao(1980)]{rao80} Rao, N.K. 1980, The Observatory, 100, 164
\bibitem[Rao et al.(1980)]{rao80a} Rao, N.K., Ashok, N.M., \& Kulkarni, P.V. 1980, JApA, 1, 71
\bibitem[Rao \& Raveendran(1993)]{raor93} Rao, N. K., \& Raveendran, A. V. 1993, A\&A, 274, 330
\bibitem[Rao et al.(1993)]{raog93} Rao, N. K., Giridhar, S., \& Lambert, D. L. 1993, A\&A, 280, 201
\bibitem[Rao(1995)]{raog95} Rao, N. K. 1995, BASI, 23, 351
\bibitem[Rao et al.(1999)]{rao99} Rao, N.K. et al. 1999, MNRAS, 310, 717
\bibitem[Rao \& Lambert(1993)]{rao93} Rao, N. K., \& Lambert, D. L. 1993, PASP, 105, 574 
\bibitem[Rao \& Lambert(2000)]{rao00} Rao, N. K., \& Lambert, D. L. 2000, MNRAS, 313, L33
\bibitem[Rao et al.(2006)]{rao06} Rao, N. K., Lambert, D. L., \& Shetrone, M. D. 2006, MNRAS, 370, 941
\bibitem[Rao \& Lambert(2008)]{rao08} Rao, N. K., \& Lambert, D. L., 2008, MNRAS, 384, 477
\bibitem[Rao \& Lambert(2010)]{rao08} Rao, N. K., \& Lambert, D. L., 2010, in Recent Advances in Spectroscopy: Theoretical, Astrophysical and Experimental Perspectives, ed. R.J. Chaudhuri, M.V. Mekkaden, A.V. Raveendran, \& A.S. Narayanan (Berlin: Springer)  177
\bibitem[Rao \& Nandy(1986)]{rao86} Rao, N. K., \& Nandy, K. 1986, MNRAS, 222, 357
\bibitem[Rosenbush(1995)]{ros95} Rosenbush, A.E. 1995, AN, 316, 213
\bibitem[Scott et al.(1997)]{sco97} Scott, A. D., Duley, W. W., \& Jahani, H. R. 1997a, ApJ, 490, L175
\bibitem[Siedel(1957)]{sie57} Siedel, Th. 1957, Mitt. Ver\"{a}nderliche Sterne, Nr. 244
\bibitem[Soker \& Clayton(1999)]{sok99} Soker, N., \& Clayton, G. C. 1999, MNRAS, 307, 993
\bibitem[Soszy\'{n}ski et al.(2009)]{sos09} Soszy\'{n}ski, I. et al. 2009, AcA, 59, 335
\bibitem[Stein et al.(1969)]{stein69} Stein, W. A., Gaustad, J., Gillett, F. C., Knacke, R. F. 1969, ApJ, 155, L3
\bibitem[Strecker(1975)]{str75} Strecker, D.W. 1975, AJ, 80, 451
\bibitem[Tenenbaum et al.(2005)]{ten05} Tenenbaum, E.D. et al.  2005, AJ, 130, 256
\bibitem[Tisserand et al.(2009)]{tiss09}Tisserand, P. et al. 2009, A\&A, 501, 985
\bibitem[Tokunaga(2000)]{tok00} Tokunaga, A.T. 2000,  Allen's Astrophysical Quantities, ed. A.N. Cox ( 4th edn. New York:Springer,)  
\bibitem[van Blerkom \& van Blerkom(1978)]{vanBl78} van Blerkom, J., \& van Blerkom, D. 1978, ApJ, 225, 482
\bibitem[Vanture et al.(1999)]{van99} Vanture, A.D., Zucker, D., \& Wallerstein, G. 1999, ApJ, 514, 932
\bibitem[Walker(1986)]{walk86} Walker, H. J. 1986, in IAU.Coll. 87,  Hydrogen Deficient Stars \& Related Objects, ed. K. Hunger, D. Sch\"{o}nberner, \& N.K. Rao (Dordrecht:Reidel) 407
\bibitem[Wdowiak(1975)]{wdo75} Wdowiak, T. J. 1975, ApJ, 198, L139
\bibitem[Werner et al.(2004)]{wer04} Werner, M. et al. 2004, ApJS, 154, 1
\bibitem[Whitney et al.(1993)]{whit93} Whitney, B. A., Balm, S. P., \& Clayton, G. C. 1993, ASPC, 45, 115
\bibitem[Woitke, Goeres \& Sedlmayr(1996)]{woitk96} Woitke, P., Goeres, A., \& Sedlmayr, E. 1996, A\&A, 313, 217
\bibitem[Yakovina et al.(2009)]{yak09} Yakovina, L.A., Pugach, A.F. \& Pavlenko, Ya.V. 2009, Astr. Reports, 53, 187
\bibitem[Za\v{c}s et al.(2007)]{zacs07} Za\v{c}s, L., Mondal, S., Chen, W. P., Pugach, A. F., Musaev, F. A., Alksnis, O. 2007, A\&A, 472, 247
\bibitem[Zaniewski et al.(2005)]{zan05} Zaniewski, A., Clayton, G.C., Welch, D.L., Gordon, K.D., Minniti, D., \& Cook, K.H. 2005, AJ, 130, 2293 
\bibitem[Zavatti(1975)]{zav75} Zavatti, F. 1975, IBVS, 1027, 1
\bibitem[Zhilyaev et al.(1978)]{zhi78} Zhilyaev, B.E., Orlov, M.Ya., Pugach, A.F., Rodriges, M.G., \& Totochava, A.G. 1978, R Coronae Borealis-type stars, (Naukova dumka: Kiev)
\end{thebibliography}
\end{document}